\documentclass[sigconf,authorversion]{acmart}

\usepackage{xcolor}

\usepackage{dirtytalk}
\usepackage{multirow}
\usepackage{float}

\AtBeginDocument{%
  \providecommand\BibTeX{{%
    \normalfont B\kern-.5em{\scshape i\kern-.25em b}\kern-.8em\TeX}}}

\copyrightyear{2023} 
\acmYear{2023} 
\setcopyright{acmlicensed}\acmConference[CHI '23]{Proceedings of the 2023 CHI Conference on Human Factors in Computing Systems}{April 23--28, 2023}{Hamburg, Germany}
\acmBooktitle{Proceedings of the 2023 CHI Conference on Human Factors in Computing Systems (CHI '23), April 23--28, 2023, Hamburg, Germany}
\acmPrice{15.00}
\acmDOI{10.1145/3544548.3581355}
\acmISBN{978-1-4503-9421-5/23/04}

\usepackage{array}
\begin{document}

\newcommand{\PreserveBackslash}[1]{\let\temp=\\#1\let\\=\temp}
\newcolumntype{C}[1]{>{\PreserveBackslash\centering}p{#1}}
\newcolumntype{R}[1]{>{\PreserveBackslash\raggedleft}p{#1}}
\newcolumntype{L}[1]{>{\PreserveBackslash\raggedright}p{#1}}

\title[Augmenting Digital Media Consumption via Critical Reflection to Increase Compassion]{``We need to do more ... I need to do more'': Augmenting Digital Media Consumption via Critical Reflection to Increase Compassion and Promote Prosocial Attitudes and Behaviors}


\author{Ken Jen Lee, Adrian Davila, Hanlin Cheng, Joslin Goh, Elizabeth Nilsen, and Edith Law}
\email{{kenjen.lee, adrian.davila, hanlin.cheng, joslin.goh, enilsen, edith.law}@uwaterloo.ca}
\affiliation{%
  \institution{University of Waterloo}
  \streetaddress{200 University Avenue West}
  \city{Waterloo}
  \state{Ontario}
  \country{Canada}
  \postcode{N2L 3G1}
}

\renewcommand{\shortauthors}{Lee, et al.}

\begin{abstract}

Much HCI research on prompting prosocial behaviors focuses on methods for increasing empathy. However, increased empathy may have unintended negative consequences. Our work offers an alternative solution that encourages critical reflection for nurturing compassion, which involves motivation and action to help others. In a between-subject experiment, participants (N=60) viewed a climate change documentary while receiving no prompts (CON), reflective prompts to focus on their emotions (RE) or surprises (RS). State compassion, critical reflection, and motivation to act or learn were measured at the end of the session (post-video) and two weeks later (follow-up). Despite participants' condition not affecting compassion, critical reflection was positively correlated with post-video state compassion. RE and RS participants demonstrated deeper reflection and reported higher motivation to learn post-video, and more prosocial behavioral changes during follow-up. RS participants reported better follow-up recall than RE participants. We conclude by discussing implications on designing technology to support compassion and longer-term critical reflection.

\end{abstract}



\maketitle

\section{Introduction}

Prosocial behaviors, broadly defined as ``any action that benefits another'' \cite{schroeder2015field}, are crucial for managing global social issues such as pandemics \cite{Jin_2021} and climate change \cite{nolan2015prosocial}.
Within the HCI research community, there has been a recent trend of employing technology to promote prosocial attitudes and behaviors by fostering empathy (e.g., \cite{10.1145/3359220,10.1145/3430524.3440644,10.1145/3477322.3477334,10.1145/3468002.3468236,baughan2021someone, barbot2020makes,Herrera_2018,roswell2020cultivating,breves2020bringing,Bae_2019}), that is, ``the ability to share someone else's feelings or experiences by imagining what it would be like to be in that person's situation'' \cite{empathycamb}. 
Often, this takes the form of simulation and storytelling using immersive technologies like virtual or augmented reality (VR/AR) \cite{gaggioli2019positive, barbot2020makes,Herrera_2018,roswell2020cultivating,breves2020bringing,Bae_2019}, or ways of designing online interactive experiences \cite{10.1145/3359220,baughan2021someone}.
For instance, Herrera et al.~\cite{Herrera_2018} built a VR perspective-taking task that simulates homelessness, and found that it encouraged more participants to sign a petition supporting affordable housing.
These research could have been partially motivated by the belief that ``empathy will save the world'' \cite{bloom2017against}, 
paired with the idea that immersive technology could be the ``ultimate empathy machine'' \cite{bevan2019behind, Herrera_2018}.

While empathy's benefits have been widely demonstrated \cite{Cohen_2012,Batson_1987,rumble2010benefits,konrath2016positive}, the promotion of empathy may lead to behaviors that are unfair, immoral or harmful to others in some circumstances \cite{breithaupt2018bad, breithaupt2019dark, bloom2017against, Martingano2020}.  As an alternative to empathy, research has shown that nurturing {\it compassion} could also promote prosocial behaviors \cite{bankard2015training, stevens2021neuroscience}. 
For instance, Leiberg et al. found that a short compassion training successfully increased helping behavior in a game designed for prosocial behavior assessment \cite{leiberg2011short}.
Such results are not surprising, since a core component of compassion is the motivation to act towards alleviating others' suffering (Table \ref{tab:compassionelem}) \cite{Strauss2016}.

Our work extends existing HCI research on compassion and prosocial behaviors in several ways. First, compared to the number of studies examining empathy (e.g., \cite{10.1145/3359220,10.1145/3430524.3440644,10.1145/3477322.3477334,10.1145/3468002.3468236}), much less attention within HCI has been paid to promoting prosocial behaviors via compassion (e.g., \cite{10.1145/3290607.3308454}); this work aims to contribute to the latter.
Second, this work explores an associated construct of compassion that has received little attention in empirical work to date, namely, critical reflection. 
Critical reflection refers to the process of reflecting on one's presuppositions \cite{mezirow1990critical}, and can be used to nurture critical thinking \cite{cranton2006understanding}.
Critical thinking, in turn, is a crucial part of compassion as it allows one to ``reason about a person's experience'' \cite{Strauss2016}, recognize their suffering \cite{Strauss2016}, and make rational decisions by understanding the outcomes of one's prosocial actions aimed at alleviating others' suffering \cite{Strauss2016, bloom2017against}.
In addition, very little empirical work has examined associations between critical reflection and state compassion (i.e., a temporary experience of compassion~\cite{fridhandler1986conceptual}), and the effectiveness of using critical reflection-based interventions to foster compassion. 

To fill these gaps, this work introduces a critical reflection process (via question asking) that is designed to foster feelings of compassion, as well as increase the motivation to learn and act for the benefit of others (i.e., prosocial attitudes and behaviors) within the context of digital media consumption. 
Since compassion consists of both affective and cognitive dimensions \cite{Strauss2016}, we investigate two types of reflection processes, one that focuses on emotions, and another that focuses on cognition (specifically, on information that is incongruent with prior understandings, i.e., surprises \cite{chen2020surprise}). 
Surprise, albeit technically an emotion \cite{ekman1992there}, is cognitive in nature and is elicited due to discrepancies between experiences and expectations \cite{hu2019our}. It is also a crucial element of curiosity \cite{markey2014curiosity}.
Importantly, both feelings \cite{boud2013promoting, hammer2003role} and curiosity \cite{dyche2011curiosity,grace2015surprise} are conducive to reflection and critical thinking.

To ground the investigation, our study focuses on climate change as the target social issue. Climate change is a global crisis that has attracted much attention and a great need for attitudinal and behavioral changes \cite{moser2016reflections}. It is also a topic that the public is knowledgeable about \cite{moser2016reflections} and thus likely to be able to reflect on their existing beliefs and understanding.  Finally, a huge amount of online climate change content have been created and consumed in recent years \cite{duran2020climate}; this provides us with a realistic scenario to explore the relationship between critical reflection, compassion and prosocial attitudes and behaviors, given popular media's potential as a material to perform critical reflection on \cite{ryan2012promoting}.

Overall, this work explores the use of critical reflection as an augmentation of climate change digital media consumption in order to nurture short- and longer-term compassion, including its final component: prosocial attitudes and behaviors. Particularly, our work:
\vspace{-\topsep}
\begin{enumerate}
    \item contributes a between-subject experiment assessing the effectiveness of using critical reflection to nurture compassion in the context of digital media consumption
    \item compares emotion- versus cognition-based approaches to critical reflection and the extent they can engender various compassion-related responses
    \item introduces an adaptation of the Sussex-Oxford Compassion for Others Scale (SOCS-O) for measuring state compassion after digital media viewing and reflection
    \item discusses technological implications on designing for longer-term reflection aimed at compassion
\end{enumerate}

\section{Related Work}

\subsection{Empathy and Compassion: Constructs and Their Distinctions}

\subsubsection{Empathy}
Empathy is often seen as having two main dimensions: cognitive empathy, which refers to an intellectual understanding of another person's emotions and perspective, and affective (or emotional) empathy, which refers to the state of ``being affected by and sharing another's emotions'' \cite{Strauss2016}. Both cognitive and affective empathy are related to a number of positive behaviors across the lifespan \cite{Cohen_2012,Batson_1987,rumble2010benefits,konrath2016positive,decety2020empathy}. However, research suggests that there may be costs to feelings of empathy.
Feeling affective empathy exposes a person to the distress and suffering felt by others, and could lead to empathic distress fatigue \cite{Hofmeyer2020}, which is ``a strong aversive and self-oriented response to the suffering of others, accompanied by the desire to withdraw from a situation, disconnect from those who are suffering, and adopting depersonalizing behaviors in order to protect oneself from excessive negative feelings'' \cite{singer2014empathy}. 
This could ``paralyze people, lead[ing] them to skewed decisions, and often spark irrational cruelty'' \cite{bloom2017against}, and result in physical and emotional exhaustion, apathy, anger, and victim blaming \cite{Vachon2015,Martingano2020,Hofmeyer2020}.

\subsubsection{Technological Support for Empathy.}
HCI researchers have investigated various ways to increase internet users' empathy while interacting with digital media. 
For instance, Taylor et al. \cite{10.1145/3359220} designed empathy nudges that appeared as participants were replying to synthetic social media posts, and found mixed results in terms of empathy enhancement.
To encourage empathy for other users, participants of Baughan et al.'s study suggested increasing the amount of details provided about other users, including their background and mood \cite{baughan2021someone}. 
Existing work also explored the design of social agents aimed at encouraging empathy and foster prosociality.
Paiva et al. \cite{10.1145/3477322.3477334} proposed a framework for building such agents that involves empathy mechanisms, modulation and responses, and discussed factors affecting their effectiveness at evoking empathy and prosociality, including agent embodiment, speech and non-verbal behaviors, and other situational details.
An example of such an agent has been studied in the storytelling context \cite{spitale2022socially}.
Besides that, Barbot and Kaufman \cite{barbot2020makes} found that going through immersive virtual reality (IVR) experiences of various types increased empathy.
However, 27.6\% of the variance in empathy was explained by UX variables, with the most prominent UX variables being illusion of body ownership and agency. Whether the IVR was empathy-evoking did not moderate this relationship, pointing to the effectiveness of VR in increasing empathy by being an immersive medium for perspective-taking, regardless of media content.
Similarly, a meta-analysis by Ventura et al.~\cite{ventura2020virtual} found IVRs to be effective at increasing perspective-taking, but not necessarily empathy.
Analyzing empathy's components separately, Martingano et al.'s meta-analysis \cite{martingano2021virtual} found IVRs to be effective at increasing emotional, but not cognitive, empathy. 
To improve the effectiveness of IVRs, they encouraged the use of more reflection prompts and allowing users to personalize their virtual experiences.

\subsubsection{Compassion}
Strauss et al. \cite{Strauss2016} proposed a comprehensive definition of compassion that contains five components (Table \ref{tab:compassionelem}).
The first component, recognizing suffering, is an awareness of others' suffering, either cognitively or through unconscious physical or affective reactions.
The second component is the understanding of common humanity, that suffering (i.e., the undergoing of pain, distress or tribulation with physical, emotional and spiritual components \cite{schulz2007patient}) is a ``shared human experience'' \cite{Strauss2016}.
The third component is ``feeling empathy for the person suffering and connecting with the distress'' \cite{Strauss2016}.
The fourth component is to tolerate difficult emotions from emotional resonance without getting overwhelmed, while remaining nonjudgmental and accepting of others.
Lastly, the fifth component is acting, or a feeling of motivation, to help.
Compassion is important to investigate within HCI since it has additional components over and above empathy, including regulating negative emotions arising from affective empathy, and a desire to act, or actions, to alleviate others' suffering.
Conceptually, compassion is broader than empathy in that i) it is ``not only felt for close others (where attachment comes into play as well), but also for those we do not know'', and that ii) it can be felt for larger targets, like humanity at large, beyond specific interpersonal encounters \cite{Strauss2016}.
Neuroimaging studies found that empathy training (focused on resonating with suffering) and compassion training (meditation-related techniques that foster feelings of benevolence and kindness) activated non-overlapping brain networks \cite{singer2014empathy,Klimecki_2013}. Aligning with Strauss's definition, while empathic training produced negative affect, feelings associated with compassionate responses were positive, other-oriented and facilitated prosocial motivation and behavior \cite{singer2014empathy,Klimecki_2013}.
These outcomes suggest that compassion may lead to resilience, instead of distress and fatigue \cite{Hofmeyer2020,Peters2014}. Such findings have garnered calls from within the HCI community for more research on technologies built to foster compassion \cite{Peters2014}.

\begin{table}[t!]
  \caption{The five compassion elements by Strauss et al. \cite{Strauss2016}.}
  \label{tab:compassionelem}
  \begin{tabular}{cp{2.8in}}
    \toprule
    & Description\\
    \midrule
    1 & Recognizing suffering \\
    2 & Understanding the universality of suffering in the human experience \\
    3 & Emotional resonance \\
    4 & Tolerating uncomfortable feelings \\
    5 & Motivation to act/acting to alleviate suffering \\
  \bottomrule
\end{tabular}
\end{table}

\subsubsection{Technological Support for Compassion.}
Even though less dualistic approaches (i.e., removing the false divide between self- vs. other-compassion) to compassion research has been called for \cite{quaglia2021compassion}, HCI research has focused predominantly on self-compassion, i.e., compassion extended towards the self~\cite{neff2011self}.
A well-known example is Vincent \cite{lee2019caring}, a chatbot that was effective at increasing self-compassion when designed to receive care from the user.
IVRs have also been explored to enhance self-compassion. 
Falconer et al. \cite{falconer2016embodying} built an IVR where patients with depression practiced delivering and receiving compassion, and measured significant increases in self-compassion. Cebolla et al. \cite{cebolla2019putting} explored the use of IVR to further enhance the positive effects of self-compassion meditation via perceived body swapping and did not find significant differences.
Some HCI work approached technologically-supported compassion with inspirations from existing compassion-based practices.
Mah et al. \cite{mah2020designing}, for instance, designed an interactive artwork inspired by Tibetan Mahayana Buddhist practices that cultivated visitors' compassion.
Through a self-observation study of practicing a Buddhist compassion cultivation technique, they proposed a framework for self-observation~\cite{mah2020understanding,mah2021towards} and explored HCI implications for supporting such meditation processes, including the balance between facilitation and taking over.
Similarly, Vacca \cite{vacca2016designing} explored how a mobile-based interactive/guided loving-kindness meditation can help integrate meditation into everyday life.
Besides that, Prabhakar \cite{10.1145/3290607.3308454} explored what new mothers and their partners consider as elements of compassionate interactions (e.g., helping with things without needing to be asked) towards designing technologies to support such interactions.


\subsubsection{Rational Compassion}
Rationality has long been discussed to be inherent to compassion.  From Buddhist perspectives, for example, compassion is both an emotional response and a response grounded ``on reason and wisdom which is embedded in an ethical framework concerned with the selfless intention of freeing others from suffering'' \cite{Strauss2016}.  Following this line of thought, Bloom proposed rational compassion, a framing of compassion that highlights its cognitive components, as wanting ``to alleviate suffering and make the world a better place ... and a rational assessment of how best to do so'' \cite{bloom2017against}.
Aspects of a rational assessment include a good understanding of an action's consequences (e.g., providing small donations to a charity might actually harm the charity since processing fees and follow-up physical mails could be more costly than the donation amount itself) and understanding and avoiding biases (e.g., ``that a child in a faraway land matters as much as our neighbour's child'') \cite{bloom2017against}.

The framing of rational compassion proposes a relationship where the strengths of logical reasoning and compassion work in unison towards human flourishing. To quote Bloom \cite{bloom2017against}, 
\begin{quote}
``\textit{While sentiments such as compassion motivate us to care about certain ends---to value others and care about doing good---we should draw on this process of impartial reasoning when figuring out how to achieve those ends.}''
\end{quote}
But how can rationality be nurtured? 
One possible way is through critical reflection, which is ``principled thinking'' that is ``impartial, consistent, and non-arbitrary'' \cite{mezirow1998critical}.
Our study does not target rational compassion per se, since research on it is still premature, e.g., there are no measurements for rational compassion specifically.
However, this study builds off of these notions by exploring the link between compassion and critical reflection.

\subsection{Levels of Reflection}\label{sec:critreflection}

Reflection is defined as ``intellectual and affective activities in which individuals engage to explore their experience in order to lead to new understanding and appreciation'' \cite{boud2013reflection}. 
Prior works have proposed methods for categorizing different reflection levels.
A notable example from adult education research is Mezirow's theory on critical reflection \cite{Kember_2008}, which categorizes reflection into three levels \cite{cranton2006understanding}. The first, non-reflective actions, refers to actions that are ``habitual or thoughtful without reflection'' \cite{lundgren2016critical}. The second level,  (non-critical) reflection, are actions that involve ``validity testing of prior learning'' \cite{mezirow1991transformative}, while the third level, critical reflection, is to reflect on presuppositions \cite{mezirow1990critical}.
A critical reflection process is one that ``requires the person to see the larger view of what is operating within his or her value system'' \cite{Kitchenham2008}, involving changes to ``deep-seated, and often unconscious, beliefs'' and the formation of new belief structures \cite{Kember_2008}.
Boud et al. proposed a model of reflection that includes elements like returning to experience, attending to feelings, association, integration, validation, appropriation and outcomes of reflection (Table \ref{tab:qual}) \cite{boud2013promoting, wong1995assessing, Kember_2008}.
Van Manen identified three levels of reflection within student learning, with the first focusing on how to achieve given objectives, the second focusing on relationships between learned principles and practice, and the third focusing on ethical and political aspects of educational discourse \cite{goodman1984reflection,van1977linking}.
Fleck and Fitzpatrick's classification of five types of reflection includes R0 Description, R1 Reflective Description, R2 Dialogic Reflection, R3 Transformative Reflection and R4 Critical Reflection \cite{fleck2010reflecting}.
Unsurprisingly, these frameworks overlap. For example, Mezirow's critical reflection aligns with Fleck and Fitzpatrick's R3 and R4, Boud's association and integration aligns with Fleck and Fitzpatrick's R2. Using these overlaps, Wong et al. \cite{wong1995assessing} analyzed nurses' reflective journals using a qualitative coding scheme that first codes data using Boud et al.'s fine-grained classifications, before mapping these classifications into Mezirow's three levels of reflection.  
Other reflection coding processes like Powell's \cite{powell1989reflective} and Sparks-Langer's \cite{sparks1990reflective} suffer from limitations noted in prior works \cite{Kember_2008}, while Kember's method \cite{Kember_2008} is designed more for formal learning settings.
Our work uses Mezirow's reflection theory (previously used in various HCI research, e.g., \cite{10.5555/1599600.1599760,10.1145/2675133.2675162,10.1145/3311350.3347192}) and Wong et al.'s coding process \cite{wong1995assessing} since they are the most appropriate given our work's context (i.e., reflection on digital content).

To encourage critical reflection, several activities could be carried out, including questioning, journal writing and experiential learning \cite{cranton2006understanding}.
To facilitate the design of critical reflection questions, Cranton provided template questions representing intersections between the six habits of mind, i.e., the different aspects of how we see ``the world based on our background, experience, culture, and personality'' \cite{cranton2006understanding}, and both i) the three types of reflection and ii) the three types of knowledge (refer to Table 2.1 and 2.2 in \cite{cranton2006understanding}).
For instance, in the context of climate change digital media consumption, a question that could be created, while targeting critical reflection (i.e., premise reflection \cite{lundgren2016critical}) for the philosophical habit of mind \cite{cranton2006understanding}, is asking what and why their views on climate change has changed after experiencing the piece of media.

\subsubsection{Technological Support for Reflection}
In their meta-analysis of technologies supporting reflection, Bentvelzen et al., identified design patterns---temporal perspectives, conversations with others, comparisons between current and ideal statuses, and discoveries of fresh perspectives \cite{bentvelzen2022revisiting}.
Fleck and Fitzpatrick \cite{fleck2010reflecting} discussed various ways that reflection could be supported technologically, e.g., using technology-supported recording for R0 and question-asking for R1, and exploration of alternative perspectives for R2.
Using Fleck and Fitzpatrick's framework, a meta-analysis of personal informatics systems by Cho et al. \cite{cho2022reflection} found that to encourage transformative reflection, future systems could use reflection prompts that are more focused on the \textit{why}'s of recorded user behaviors. 

Critical reflection can also be fostered through a shared sense of community. 
Halbert and Nathan \cite{10.1145/2675133.2675162} investigated how group-based critical reflection could be supported by engaging with feelings of discomfort, and among other suggestions, proposed that such technologies should foster a shared identity and respectful dialogue, prioritize the voices and experiences of context-specific marginalized users, and support pluralism.
Citing tools built by Kriplean et al. like ConsiderIt \cite{kriplean2012supporting} and Reflect \cite{kriplean2012you}, the importance of interfaces that encourages the finding of common ground was also discussed \cite{10.1145/2675133.2675162}.
Investigating social-emotional learning contexts, Slovak et al. \cite{slovak2017reflective} proposed technological characteristics for supporting transformative reflection, including a balance of how real technologically-simulated situations are between being conducive to learning without being overly emotional, and having the capabilities to support explicit, social and personal practicum components.

\subsection{The Use and Role of Digital Media}
Digital media has become a routine part of many people's lives. Video sharing platform YouTube, from which the climate change video used in this work originated, has 1.7 billion unique monthly users who are on YouTube for a daily average of 19 minutes, streaming 694,000 hours of videos every minute \cite{youtubestats}.
Despite the enormous amount of digital media on human suffering available and shared every day, the expected translation from media consumption to compassionate behaviors towards removing these sufferings has largely not been realized, especially for distant human suffering \cite{Hoskins_2020}. 
A commonly given reason for this is an empathic distress fatigue experienced by digital media consumers due to an overload of, and the resulting over-familiarity and sense of numbness with, digital media depicting human suffering~\cite{Hoskins_2020,moeller2002compassion,ow2014our,varma2019empathy}.
Hoskins, however, provided an alternative perspective on this issue, arguing that in the context of war-related digital media, a post-truth society, where ``objective facts are less influential in shaping public opinion than appeals to emotion and personal belief'' \cite{posttruthdefinition}, has led to 
war's ``in/visibility'', an illusory transparent nature of digital media on war 
\cite{Hoskins_2020}. According to Hoskins, the capacity for digital coverage of human suffering to mobilize global action is often overestimated \cite{Hoskins_2020}.
However, this ineffectiveness to motivate prosociality might not be due to inherent characteristics of digital media, but \textit{how} digital media content is framed.


Climate change is an example of a global issue that has garnered much attention from the public through the abundance of related digital media \cite{mavrodieva2019role}. Putting aside climate change skepticism \cite{Hornsey_2019} and misinformation \cite{van2017inoculating}, which are beyond the scope of this work, 
public climate change concerns have not been translated into sufficient actions (e.g., individual actions, public activism)  to address the issue \cite{Hornsey_2019,van2017inoculating}.
While discussing which types of actions are most effective or appropriate on an individual level to mitigate climate change is beyond this work's purpose (refer to \cite{fragniere2016climate,Rickard_2014,Kenis_2012,whitmarsh2012introduction}), our study investigates how to augment the digital media consumption process to foster compassion and one of its key components: prosocial beliefs and behaviors.
The present study is motivated by prior work showing that a purely empathic response towards digital media related social and global issues is often insufficient to motivate prosocial attitudes and behaviors \cite{varma2019empathy}, and the observation/premise that not all digital media are necessarily framed in ways that effectively encourage prosociality. 
Specifically, critical reflection could motivate compassion in three ways: i) by aiding in the search of common ground and humanity, ii) complementing emotional responses with honest cognitive explorations of the relationships between a piece of digital media and one's own values and belief structure (which could also be beneficial for navigating an "in/visible" \cite{Hoskins_2020} digital world), and finally, by iii) providing a space to investigate how changes to one's own values and beliefs could transform into changes in prosocial attitudes and actions.

\section{Research Questions}

In our work, participants watched a climate change documentary, with those in reflective conditions given reflective prompts to encourage critical reflection. Control (CON) condition participants received no reflective prompts, while Reflect-Emotion (RE) and Reflect-Surprise (RS) participants received emotion-focused and cognitive-focused prompts, respectively. Participants were measured for their level of reflection, compassion, prosocial attitudes and behaviors both at the end of the session and two weeks later. Findings are aimed at informing how technology can better support critical reflection for compassion. Specifically, our work investigates the following research questions (RQ):

\begin{itemize}
\item\textbf{RQ1} What are the effects of the reflective prompts on compassion, prosocial attitudes and behaviors?\\
\textbf{H1} Participants given reflective prompts (RE \& RS) will demonstrate deeper reflection and report higher state compassion, concern, motivation to learn about what they can do about climate change (i.e., prosocial attitude) and adopt more climate change mitigation-related behaviors (i.e., prosocial behaviors).
\item\textbf{RQ2} What are the differences between an emotion-focused (RE) versus a surprise-focused (RS) reflection process?\\
\textbf{H2} RE and RS participants will focus on different (i.e., emotional vs. informational) parts of the video to reflect on.
\item\textbf{RQ3} How is compassion, regardless of condition, related to media consumption experience, reflection levels, prosocial attitudes and behaviors?\\
\textbf{H3.1} Critical reflection and both trait and state compassion are positively correlated.\\
\textbf{H3.2} State compassion is positively related to concern and motivation.\\
\textbf{H3.3} State compassion is negatively related to distress.
\end{itemize}

\section{Methodology}

\subsection{Study Design}

\begin{figure*}
    \includegraphics[width=1\textwidth]{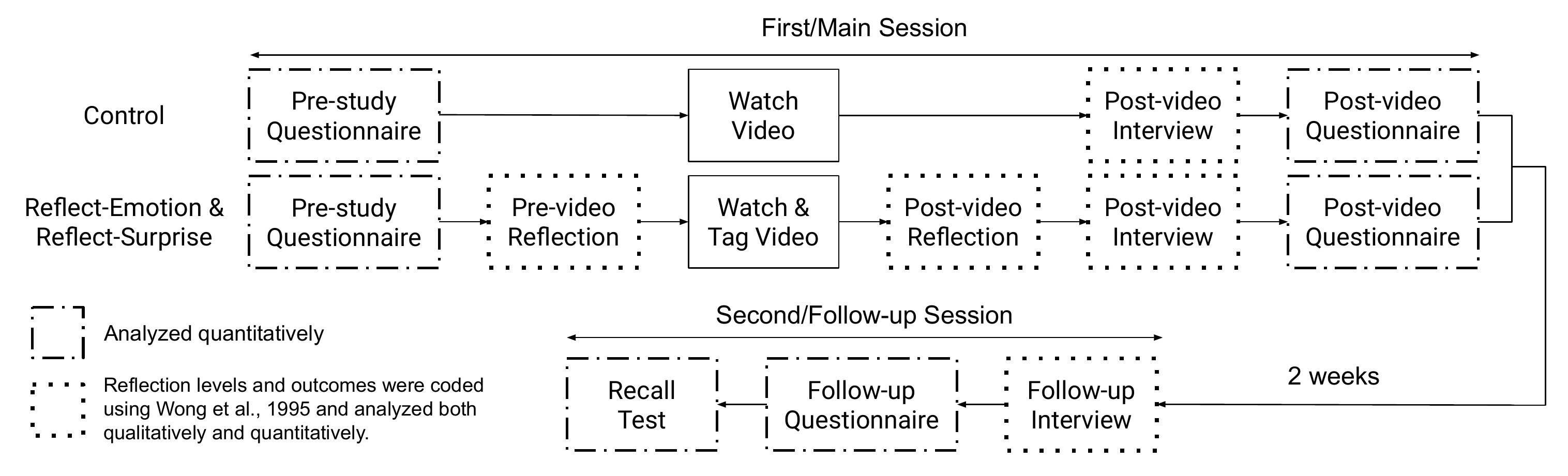}
    \caption{The flow of the study for all three conditions. Reflection questions were different for the Reflect-Emotion and Reflect-Surprise conditions (Table \ref{tab:reflectqns}).}
    \label{fig:studyflow}
    \Description{The flow of the study for all three conditions is shown in the figure. The study is split into two sessions, with the second session happening two weeks after the first session. In the first, or main, session, control participants first undergo a pre-study questionnaire. They then watched the climate change video, went through the post-video interview and lastly, the post-video questionnaire. For the Reflect-Emotion and Reflect-Surprise participants, they first went through the pre-study questionnaire. They were then asked the pre-video reflection questions, watched and tagged the video, before being asked the post-video reflection questions. The reflection questions were different for the Reflect-Emotion and Reflect-Surprise conditions. Please refer to Table 2 for the reflection questions used. Lastly, they went through the post-video interview and questionnaire. 
    For the second, or follow-up, session, all participants went through the same process, namely a follow-up interview, questionnaire, and finally a recall test.}
\end{figure*}

The study was an online, between-subject experiment.  Consent was provided electronically, questionnaires were administered via Qualtrics, instructions were provided verbally, and open-ended questions were asked through interviews via Microsoft Teams meetings.

Participants were asked to watch a 17-minute YouTube documentary created by the 60 Minutes Australia program on the Solomon Islands' rising sea levels \cite{risingseavideo}.
The video was chosen to reflect a typical type of publicly accessible climate change documentary video with an urgent tone for action \cite{o2009fear}. Its length was long enough for a meaningful presentation of both emotions and information on the topic, but not too long as to cause participant fatigue. 
Climate change issues are of high public interest \cite{dubovi2021interactions} and hence improves the study's ecological validity.

Our main goal is to investigate the use of reflection prompts in digital media consumption and their effects on compassion, prosocial attitudes and behaviors.  Participants were randomly assigned one of three conditions: Control (CON), Reflect-Emotion (RE) or Reflect-Surprise (RS). 
The conditions' procedures are illustrated in Figure \ref{fig:studyflow}; CON does not include reflection questions, RE includes pre-video and post-video reflection questions intended to elicit reflection on emotions, while RS includes questions intended to elicit reflection on surprising information.

Specifically, for CON participants, the first session consists of a pre-study questionnaire, video-viewing session, a post-video interview and questionnaire, and averaged 39 minutes 27 seconds in duration.  In contrast, after the pre-study questionnaire and before viewing the video, RE and RS participants were asked reflection questions about their values and beliefs (pre-video in Table \ref{tab:reflectqns}). 
This is important, since a clear view of one's existing beliefs, experiences and knowledge is crucial to fully engage in reflection \cite{nolan2008encouraging}.
RE/RS participants were then asked to install the \textit{TimeTags for YouTube} extension on Google Chrome \cite{timetag}, and used it to tag (i.e., bookmark) moments in the video based on different instructions: 

\begin{quote}
    RE: ``You will now watch a video on rising sea levels. Please tag moments in the video that make you feel either positive or negative emotions.''
\end{quote}
\begin{quote}
    RS: ``You will now watch a video on rising sea levels. Please tag moments in the video that make you feel surprised.''
\end{quote}

To familiarize themselves with the tagging procedure, RE and RS participants were given the opportunity to practice by tagging a test video \footnote{The test video used was a video titled ``White Noise 3 Hour Long'' on YouTube.} using the extension.

After watching the climate change video, RE and RS participants were asked to select two tags that were either ``the most emotional, positive or negative'' (for RE), or ``the most surprising'' (for RS).
For each chosen tag, a set of reflection questions were asked (post-video in Table \ref{tab:reflectqns}). 
The use of specific moments from the video as targets of reflection helped make the reflection process more focused and detailed, both of which are important characteristics of initial reflection stages \cite{boud2013promoting}.
Participants only reflected on two tags to prevent cognitive fatigue, which could affect reflection quality \cite{timmons2019moral}.
This method of reflecting on a past experience (i.e., watching the video) through tagging aligns with    temporal perspective reflection methods identified by Bentvelzen et al. \cite{bentvelzen2022revisiting}.
After that, three higher-level reflection questions were also asked (post-video in Table \ref{tab:reflectqns}).
These questions aimed to guide participants through the components of compassion by critically reflecting on lessons learned from the tags and how those lessons could convert into prosocial attitudinal and behavioral changes. 
Specifically, the RE condition goes beyond encouraging empathy and targets compassion through reflections on the deeper reasons and impacts of empathic feelings felt by participants.
Lastly, RE and RS participants were given the post-video questionnaire, concluding the first session. On average, the first session lasted 58 minutes 13 seconds for RE and RS participants.

Two weeks after the first session, all participants (CON, RE and RS) attended a follow-up session (lasting an average of 11 minutes 4 seconds) that assessed longer-term effects of the first session in terms of their reflection, feelings and recollection about the video. 

\begin{table*}
  \caption{Reflection questions used in the Reflect-Emotion (RE) and Reflect-Surprise (RS) conditions, created based on Cranton's framework for nurturing critical reflection \cite{cranton2006understanding}.}
  \label{tab:reflectqns}
  \begin{tabular}{ccp{4.5in}}
    \toprule
    Time & Condition & Question\\
    \midrule
    \multirow{6}{*}{Pre-video} & \multirow{6}{*}{RE \& RS} & Please describe your views and values on climate change and the rising sea levels. \\
    && Please describe your views on taking actions to mitigate climate change and its impacts. \\
    && Please describe your views on your role in taking actions to mitigate climate change and its impacts. \\
    && How did you come to your views and values described above, and how, if any, have they changed over time? \\
    \midrule
    \multirow{16}{*}{Post-video} & \multirow{8}{*}{RE} & [For each chosen tag] What did you feel at this moment, and what at this moment caused you to feel how you felt? \\
    && [For each chosen tag] How did your experience and values influence how you felt at this moment? \\
    && [For each chosen tag] How did your views and beliefs change after this moment? \\
    && How have your views and values on climate change and the rising sea levels changed? \\
    && Why have they changed in the way described above? \\
    && How will this change influence you and your behavior going forward? \\
    \cline{2-3}
    & \multirow{8}{*}{RS} & [For each chosen tag] What is the information presented at this moment? \\
    && [For each chosen tag] What were your initial understanding and assumptions about the information presented at this moment? \\
    && [For each chosen tag] How did your understanding and assumptions change after this moment? \\
    && How have your views and values on climate change and the rising sea levels changed? \\
    && Why have they changed in the way described above? \\
    && How will this change influence you and your behavior going forward? \\
  \bottomrule
\end{tabular}
\end{table*}

\subsection{Measures}
\subsubsection{Pre-Study Measures}
These measures were administered as part of the pre-study questionnaire (Fig. \ref{fig:studyflow}).

\noindent\textbf{Demographic questions.} Participants were asked about their demographic, including their age, ethno-racial background(s), gender identity(ies), educational attainment level, university education background (STEM or BHASE) and whether they speak English as a first language (Table \ref{tab:participants}).

\noindent\textbf{Sussex-Oxford Compassion for Others Scale (SOCS-O).} The SOCS-O is a 20-item scale where each question is rated on a 5-point Likert scale from 1:``Not at all true'' to 5:``Always true''. 
The scale consists of five subscales that measure the five dimensions of compassion, namely, (i) recognizing suffering, (ii) understanding the universality of suffering, (iii) feeling for the person suffering, (iv) tolerating uncomfortable feelings, and (v) motivation to act, or acting, to alleviate suffering \cite{Gu2019}. 
Participants' total score on this scale (/$100$) were used for analyses.

\noindent\textbf{Short Form Marlowe-Crowne Social Desirability Scale (M-C SDS).} Social desirability bias has been found to be a confounding factor affecting prosocial attitudes and behaviors, especially in the context of a research study \cite{eisenberg1990empathy,runyan2019using}.  This is a 10-item binary (0:false; 1:true) scale measuring social desirability, which is ``a need for social approval and acceptance and the belief that this can be attained by means of culturally acceptable and appropriate behaviors'' \cite{marlow1961social}. According to a meta analysis of the various short forms of SDS, Fischer and Fick \cite{fischer1993measuring} found that the 10-item version by Strahan and Gerbasi \cite{strahan1972short} provides the best measure of social desirability. The score is out of $10$; a higher score indicated greater need for social approval.

\noindent\textbf{Attitudes towards climate change.} 
Three 5-point Likert scale (1-5) questions from \cite{Howell2011} were included, asking participants about their level of concern about climate change (from 1:``Not at all concerned'' to 5:``Very concerned''), motivation to try to do something about climate change (referred to as motivation to act; from 1:``Not at all true'' to 5:``Extremely true''), and motivation to learn about what they can do about climate change (referred to as motivation to learn; from 1:``Not at all true'' to 5:``Extremely true'').

\noindent\textbf{How emotional as a person.} 
This 5-point Likert scale question (from 1:``Not emotional at all'' to 5:``Extremely emotional'') asking how emotional participants perceived themselves as was included, since it may affect how they experience emotional aspects of the documentary.

\noindent\textbf{How curious as a person.} 
Similarly, a 5-point Likert scale question (from 1:``Not curious at all'' to 5:``Extremely curious'') asking how curious participants perceived themselves as was included, since it may affect how they experience informational aspects of the documentary.

\subsubsection{Post-Study Measures}\label{sec:time1measure} 
These measures were administered as part of the post-video questionnaire and interview (Fig. \ref{fig:studyflow}).

\noindent\textbf{State personal distress.} This was measured using a 7-point Likert scale (from 1:``Not at all'' to 7:``Extremely'') measuring how strongly participants feel each given emotion, which are \textit{alarmed, grieved, upset, worried, disturbed, perturbed, distressed,} and {\it troubled} \cite{Batson1987}. The total score (out of 56) was used in the analysis. 

\noindent\textbf{State compassion.} 
Existing state compassion measurements, to the best of the authors' knowledge, do not fully measure the five elements of compassion (Table \ref{tab:compassionelem}). For example, \cite{piff2010having} used two items that only measured suffering recognition and motivation to care for others who are vulnerable.
As such, we created a questionnaire (Table \ref{tab:statecomp}), adapted from the SOCS-O (which measures \textit{trait} compassion), to measure participants' state compassion.  The questionnaire consists of eight items rated on a 5-point Likert scale (from 1:``Not at all true'' to 5:``Extremely true''). These eight items were chosen, out of the SOCS-O's 20 items, due to their suitability in measuring state in this study's context; for example, an item like ``I recognize when other people are feeling distressed without them having to tell me'' is not suitable since the group discussed in the video (the Solomon Islanders) did not interact directly with participants.

\begin{table*}
  \caption{The eight items used to measure state compassion adapted from SOCS-O and its subscales \cite{Gu2019}. Each item is rated on a 5-point Likert scale from ``Not at all true'' to ``Extremely true''. The terms ``their'' or ``them'' refer to the main subjects of the chosen video in this study: the people living in the Solomon Islands.}
  \label{tab:statecomp}
  \begin{tabular}{clp{3in}}
    \toprule
    No. & Subscale & Item Description\\
    \midrule
    1 &Recognizing suffering& I recognize their suffering and distress. \\
    2 &Understanding the universality of suffering& Like me, they experience struggles in life. \\
    3 &Feeling for the person suffering& I feel concerned for their wellbeing. \\
    4 &Feeling for the person suffering& I feel their distress. \\
    5 &Tolerating uncomfortable feelings& I am open to their feelings, instead of avoiding them. \\
    6 &Tolerating uncomfortable feelings& I connect with their suffering without judgements. \\
    7 &Tolerating uncomfortable feelings& I do not feel overwhelmed by their distress. \\
    8 &Motivation to act, or acting, to alleviate suffering& I want to try to do things that might be helpful for them. \\
  \bottomrule
\end{tabular}
\end{table*}

\noindent\textbf{Equality of value of lives.} A question was created to assess a key characteristic of rational compassion; i.e., one's ability to understand and avoid biases (e.g., “that a child in a faraway land matters as much as our neighbour’s child”) \cite{bloom2017against}. It asked participants, on a 5-point Likert scale from 1:``Not at all true'' to 5:``Extremely true'', ``How true is it that the lives of the people living in the Solomon Islands have the same value as the lives of those I love.''

\noindent\textbf{Reflective-Thinking Scale (RTS).} This contains items, rated on a 5-Likert scale from 1:``Definitely disagree'' to 5:``Definitely agree'', that measure the level of reflective thinking \cite{Kember2000,Zhang2018}. The original scale has 16 items that are evenly split into four subscales.
Only the Reflection and Critical Reflection subscales were administered, as the other subscales were less relevant for our research purposes \cite{Zhang2018}.
Each subscale score was calculated as the mean score of the subscale's items.
Two sets of RTS were administered to examine participants' reflective processes elicited as a result of the video (RTS Vid) and the verbal/reflection questions asked by the interviewer before/after the video (RTS Qn).

\noindent\textbf{Changes in attitudes towards climate change.}
Five-point Likert-scale questions (from 1:``Reduced significantly'' to 5:``Increased significantly''), adapted from \cite{Howell2011}, on how participants' i) level of concern, ii) motivation to act, and iii) motivation to learn changed from the pre-study survey to the post-video survey. Assessing attitude changes allows us to account for potential ceiling effects, where certain participants may have an extremely high level of initial concern or motivation at the onset of the study \cite{Howell2011}.

\noindent\textbf{Open-ended questions.} Two open-ended questions were adapted from \cite{Howell2011} to understand if participants had reflected during the session. The first asks participants ``What message are you taking away from the video?'' while the second asks participants ``Whose responsibility is it to work on the reasons behind rising sea levels?'' Both questions were asked during the post-video interview.

\subsubsection{Follow-up Measures}\label{sec:time2measure} To examine whether there was a longer term impact from the conditions, participants were administered an interview, questionnaire and recall test two weeks after the first session (follow-up session in Fig. \ref{fig:studyflow}).  
Post-video measures that were re-administered  included state personal distress, state compassion, equality of value of lives, the Reflective-Thinking Scale, changes in attitudes related to climate change and the two open-ended questions. Additionally, these following measures were included:

\noindent\textbf{Recall test.} Six multiple-choice questions were created based on the video's content, aimed at measuring how well participants remembered the topics discussed in the video. Recall also serves as a way to measure participants' engagement with the video's information.
Each question was given a score of one if answered correctly and zero otherwise, for a total possible score of six.

\noindent\textbf{Additional open-ended questions.} Three additional questions were asked during the follow-up interview to understand changes, if any, in participants' behaviors and attitudes in the two weeks after the first session:
\vspace{-1mm}
\begin{enumerate}
    \item “Were there times when you thought of the study in the past two weeks? If so, can you describe when they were?”
    \item “Have you taken any initiatives to learn more about climate change in the past two weeks?”
    \item “Have you done anything new related to climate change in the past two weeks?”
\end{enumerate}

\subsection{Coding of Reflection and Interview Responses}\label{sec:qualcoding}

\begin{table*}[h]
\caption{Descriptions of the first-level qualitative coding scheme based on Wong et al.'s methodology \cite{wong1995assessing}.}
\label{tab:qual}
\begingroup
\small
\begin{center}
\begin{tabular}{ p{0.05\linewidth} c c }
  Index & Name & Definition \& Guiding Questions (GD)  \\
  \hline
  BR1 & Returning to Experience & \multirow{5}{4.2in}{a) Recollecting what has taken place and replaying the experience b) To observe the event as it has happened and to notice exactly what occurred and one's reactions to it in all its elements.\\GD1. Do they describe an event that shapes their reflection/beliefs on climate change?\\GD2. Do they go beyond providing factual information or superficial descriptions of their reactions?}  \\ \\ \\ \\ \\
  \hline
  BR2 & Attending to Feelings &  \multirow{4}{4.2in}{Utilizing positive feelings via retainment and enhancement, and recognizing and removing obstructing feelings. Note that this is not just purely stating how one feels.\\GD1. Are they reflecting on their positive feelings and not just making a passing comment?\\GD2. Are they going beyond BR1 and diving deeper into their emotions?}  \\ \\ \\ \\ 
  \hline
  BR3 & Association/Integration & \multirow{10}{4.2in}{a) Linking prior knowledge, feelings or attitudes with new knowledge, feelings or attitudes b) Discovering prior knowledge, feelings or attitudes that are no longer consistent with new knowledge, feelings or attitudes c) Re-assessing prior knowledge, feelings or attitudes and modify to accommodate new knowledge, feelings or attitudes. d) Seeking the nature of relationships of prior knowledge, feelings or attitudes with new knowledge, feelings or attitudes e) Drawing conclusions and arriving at insights into the video.\\GD1. Is the new knowledge, feeling or attitude gained/elicited from the video?\\GD2. Is there an explicit connection to prior knowledge, feelings or attitudes?\\GD3. Does the participant go beyond the link between new and old, and expand on either what the connection implies, or other insights gained from the connection?}  \\ \\ \\ \\ \\ \\ \\ \\ \\ \\
  \hline
  BR4 & Validation &  \multirow{6}{4.2in}{Testing for internal consistency between new appreciations and prior knowledge or beliefs, and trying out new perceptions in new situations.\\GD1. Does the participant demonstrate both a new appreciation and an application in new situations?\\GD2. Is the participant experimenting with the new perception with an attitude of trying to assess it as an external entity, instead of as an expression of something they belief in? }  \\ \\ \\ \\ \\ \\ 
  \hline
  BR5 & Appropriation & \multirow{5}{4.2in}{a) Making knowledge one’s own b) New knowledge, feelings or attitude entering into one's sense of identity or becoming a significant force in one's life\\GD1. Does the participant express the new knowledge, feeling or attitude as their own?\\GD2. Is the participant making the knowledge their own via a process that makes the knowledge important personally? }  \\ \\ \\ \\ \\ 
  \hline
  BR6 & Outcome of Reflection &  \multirow{13}{4.2in}{a) Transformation of perspectives b) Change in behavior c) Readiness for application d) Commitment to action\\Do you see any of:\\GD1. A transformation of perspectives: Did the participant undergo a significant shift in how they look at this issue?\\GD2. A change in behavior: Did the participant do something beyond what they have already been doing since before the study? Is this behavior non-trivial (e.g., beyond just describing their participation in the study to others)? Did the participant take any initiatives in terms of outward actions to learn more about the issue due to the first session?\\GD3. Readiness for application: Did the participant demonstrate an intention to help with the issue, or apply the new knowledge, feelings or attitudes in new ways?\\GD4. A commitment to action: Did the participant promise that they would personally carry out certain actions inspired by their new knowledge, feelings or attitudes? }  \\ \\ \\ \\ \\ \\ \\ \\ \\ \\ \\ \\ \\ 
 \hline
\end{tabular}
\end{center}
\endgroup
\end{table*}

Responses to all reflection (Table \ref{tab:reflectqns}) and interview (Sections \ref{sec:time1measure} and \ref{sec:time2measure}) questions were coded based on Wong et al.'s methodology \cite{wong1995assessing}.
In this process, qualitative responses are first coded based on the elements of Boud et al.'s reflection model \cite{boud2013promoting}, including Returning to Experience, Attending to Feelings, Association, Integration, Validation, Appropriation and Outcome of Reflection (Table \ref{tab:qual}).
Following that, in the second level of coding, each participant is categorized into their Mezirow reflection classification; either a non-reflector, a reflector, or a critical reflector. 
A critical reflector demonstrates any of Validation, Appropriation or Outcome of Reflection. Otherwise, if any of Returning to Experience, Attending to Feelings, Association, or Integration is/are demonstrated instead, they are categorized as a reflector.
Lastly, if none of Boud's reflection elements is demonstrated, they are non-reflectors \cite{wong1995assessing}.
In terms of inter-coder reliability (ICR), Wong et al. used Miles and Huberman's percentage-based calculation \cite{miles1994qualitative}, which does not account for agreement due to chance \cite{O_Connor_2020}. 
Instead, following O’Connor and Joffe's \cite{O_Connor_2020} recommendation, we used Krippendorff's Alpha \cite{krippendorff2011computing}.  
Wong et al.'s methodology \cite{wong1995assessing} was used in this work as following.
One participant from each condition was randomly chosen and coded independently by the first three authors based on Wong et al.'s coding scheme \cite{wong1995assessing}.
The initial round of coding resulted in a low first-level ICR ($M_\alpha = .11\pm.32$) and an acceptable second-level ICR ($\alpha = .95$).
A subsequent group discussion revealed individual coding differences due to subjective interpretations of the codes, resulting in several changes: a) Association and Integration were merged into a single code, reflecting the difficulty in clearly separating between the two elements, an issue also faced by prior works \cite{wong1995assessing,spencer1999use}; b) definitions of the codes were made more specific to the study's context and guiding questions were added. The final coding scheme is available in Table \ref{tab:qual}.
Following this, independent coding of a second set of three randomly chosen participants (one from each condition) yielded an acceptable first-level ICR of $M_\alpha = .78\pm.42$ (primarily due to slight confusions on BR6) and perfect second-level ICR ($\alpha = 1$).
Finally, the remaining data was divided among the first three authors for independent coding. 
Coders determined participants' second level codes separately, once for the first session (i.e., session one Mezirow reflection classification), and another for the second session (i.e., session two Mezirow reflection classification).
Moreover, a separate binary code was created for prosocial behavioral changes in the second session (behavioral change is a subset of BR6; Table \ref{tab:qual}).
These codes were used in the quantitative analysis (Sections \ref{sec:result_rq1analysis}, \ref{sec:result_rq2analysis} and \ref{sec:result_rq3analysis}).

\begin{table*}[t!]
\caption{Descriptive statistics of participants. 
Participants' ethno-racial backgrounds' simple effects were insignificant ($\chi^2 (14, N= 60)$ = 10.11, \textit{p} = .75) and presented in Appendix \ref{appendix:ethnoracial} for brevity.
}
\label{tab:participants}
\begingroup
\begin{center}
\begin{tabular}{ p{0.17\linewidth} c c c c }
   & Control (\textit{n}=20) & Reflect-Emotion (\textit{n}=20) & Reflect-Surprise (\textit{n}=20) &  \\
  \hline
 Age (years) & \textit{M}=24.75$\pm5.95$ & \textit{M}=24.2$\pm3.78$ & \textit{M}=23.3$\pm3.96$ & \textit{F}(2,57)=0.49, \textit{p} = .61  \\  
 \hline
 Gender  & 5 man & 8 man & 7 man & $\chi^2 (4, N= 60)$ = 2.85 \\
  & 14 woman & 12 woman & 13 woman & \textit{p} = .58 \\
  & 1 agender \&  &  & & \\
  & genderqueer &  & & \\
 \hline
 Degree completion$^1$ & 10 completing & 9 completing & 9 completing & $\chi^2 (2, N= 60)$ = .13\\ 
  & 10 completed & 11 completed & 11 completed & \textit{p} = .94  \\
 \hline
 STEM  & 16 yes, 4 no & 13 yes, 7 no & 14 yes, 6 no & $\chi^2 (2, N= 60)$ = 1.15, \\
   & & & & \textit{p} = .56  \\
 \hline
 English as first  & 9 yes, 11 no & 9 yes, 11 no & 6 yes, 14 no & $\chi^2 (2, N= 60)$ = 1.25, \\
 language &  &  &  & \textit{p} = .54  \\
 \hline
 How emotional as a   & \textit{M}=3.55$\pm0.83$ & \textit{M}=3.35$\pm0.81$ & \textit{M}=3.3$\pm0.66$ & $\chi^2 (2, N= 60)$ = 1.07,\\
  person (1-5) & & & &  $p = .58$ \\
  \hline
 How curious as a   & \textit{M}=3.7$\pm0.66$ & \textit{M}=3.75$\pm0.97$ & \textit{M}=3.7$\pm0.57$ & $\chi^2 (2, N= 60)$ = .10,\\
 person (1-5) &  &  & & \textit{p} = .95 \\
 \hline
 Concern in climate   & \textit{M}=4.25$\pm1.02$ & \textit{M}=4.15$\pm0.93$ & \textit{M}=4$\pm0.65$ & $\chi^2 (2, N= 60)$ = 2.32,\\
 change (1-5)&  & &  & \textit{p} = .31 \\
 \hline
 Motivation to act  & \textit{M}=3.3$\pm0.92$ & \textit{M}=3.5$\pm1$ & \textit{M}=3.45$\pm1$ & $\chi^2 (2, N= 60)$ = .23,\\
 (1-5)&  &  &  & \textit{p} = .89 \\
 \hline
Motivation to learn  & \textit{M}=3.7$\pm1.03$ & \textit{M}=3.75$\pm1.33$ & \textit{M}=3.6$\pm.99$ & $\chi^2 (2, N= 60)$ = .77,\\
(1-5)&  & &  & \textit{p} = .68 \\
 \hline
 SOCS-O & \textit{M}=78.74$\pm10.14$ & \textit{M}=78.85$\pm9.77$ & \textit{M}=79.47$\pm6.71$ & \textit{F}(2,55)=0.04, \textit{p} = .96  \\  
 \hline
 SDS & \textit{M}=4.75$\pm2.10$ & \textit{M}=4.7$\pm1.84$ & \textit{M}=4.9$\pm2.20$ & \textit{F}(2,57)=0.05, \textit{p} = .95  \\  
 \hline
 Prior knowledge  & \textit{M}=2.15$\pm.99$ & \textit{M}=1.3$\pm.66$ & \textit{M}=1.75$\pm.91$ & $\chi^2 (2, N= 60)$ = 9.58, \\
 (1-5)$^2$ && & &\textit{p} = .01 \\
 \hline
 Trust in video  & \textit{M}=4.25$\pm.64$ & \textit{M}=4.4$\pm.68$ & \textit{M}=4.4$\pm.60$ & $\chi^2 (2, N= 60)$ = .82, \\
 (1-5)$^2$ &&  &&\textit{p} = .66 \\
 \hline
\end{tabular}
\end{center}
\endgroup
\begin{flushleft}
\textit{$^1$Note: This is based on the Canadian census question on an individual's highest educational attainment \cite{canadacensus}. All participants had either ``Some postsecondary education'' or had attained a ``Postsecondary certificate, diploma or degree''.}\\
\textit{$^2$Note: These two questions were asked in the post-video survey after participants watched the video.}
\end{flushleft}
\end{table*}

\subsection{Participants}\label{sec:participants}

A total of 68 participants were recruited via mailing lists, social media, and physical posters in the authors' university.
Participants received a C\$10 Amazon gift card for each of the two sessions.  
The data of the first five participants was treated as pilot data and not included in the analysis. Among the other 63 participants, three were removed from analysis due to experimenter error (n=1), technical issues (n=1), and 
a lack of willingness to engage in the reflection (n=1, assigned the RE condition but had zero tags to reflect on because they felt no particular emotions during video-watching, since ``the video could have been engineered ... [through the] editing'').
As such, data from 60 participants were analyzed. There were 20 participants in each of the three conditions, labelled as CON1 to CON20, RE1 to RE20, and RS1 to RS20. Demographics information is summarized in Table \ref{tab:participants}.

\section{Results}

\subsection{Validation and Reliability of the Adapted SOCS-O as a Measure of State Compassion}

Since participants' state compassion is an outcome measure for all three RQs, we first conducted a basic psychometric analysis to verify the reliability and validity of the adapted SOCS-O \cite{Gu2019}, which was used to measure participants' state, instead of trait, compassion on an eight-item scale (Table \ref{tab:statecomp}).

\subsubsection{Reliability}
The adapted SOCS-O scale was found to have acceptable reliability in this sample, with Cronbach’s alpha values \cite{bland1997statistics} of $\alpha = .80$ at post-video and $\alpha = .79$ at follow-up.

\subsubsection{Validity}

\begin{table*}[htbp!]
\caption{Correlation of post-video state compassion with relevant measurements.}
\vspace{-5mm}
\begingroup
\begin{center}
\label{tab:compcor}
\begin{tabular}{ p{0.6\linewidth} c c }
  & Pearson's $r$ & $p$ \\
  \hline
  (Pre-study) SOCS-O & .48 & $<.001$\\
  (Pre-study) SOCS-O - Recognizing suffering & .13 & $.32$\\
  (Pre-study) SOCS-O - Understanding the universality of suffering & .43 & $<.001$\\
  (Pre-study) SOCS-O - Feeling for the person suffering & .44 & $<.001$\\
  (Pre-study) SOCS-O - Tolerating uncomfortable feelings & .37 & $.004$\\
  (Pre-study) SOCS-O - Motivation to Act/Acting to alleviate suffering & .47 & $<.001$\\
  (Post-video) Equality of value of lives & .53 & $<.001$\\
  (Follow-up) State compassion & .28 & $.03$\\
  \hline
\end{tabular}
\end{center}
\endgroup
\end{table*}

As shown in Table \ref{tab:compcor}, regardless of condition, participants' post-video state compassion was significantly and positively correlated with all but one subscale of the full SOCS-O questionnaire. The similarity of post-video state compassion to pre-study trait compassion was expected, since the time difference between pre- and post-video questionnaires is considerably short. Post-video state compassion was also significantly correlated to how equal participants think the value of the Solomon Islanders' lives are when compared to the value of their loved ones' lives at post-video.
However, post-video state compassion was not correlated with follow-up state compassion two weeks later; this was expected, since the two weeks between the two sessions was sufficiently long for participants' state compassion to have changed significantly \cite{chaplin1988conceptions}. 
Just like at post-video, follow-up state compassion and follow-up equality of value of lives were significantly correlated ($r = .49, p < .001$).

\subsection{RQ1: Effects of Reflection Process}\label{sec:result_resproc}

\subsubsection{Analysis Methods}\label{sec:result_rq1analysis} 

Regression analysis was used to explore RQ1. 
In particular, a model was built for each post-video and follow-up measurement (i.e., Sections \ref{sec:time1measure},  \ref{sec:time2measure}, and codes from Section \ref{sec:qualcoding}) as the outcome variable.
In order to analyze the effect of the reflection process, participants' condition was coded as 0 for CON participants, and 1 for RE or RS participants (i.e., RE and RS participants are analyzed together as a group).
To control for individual differences, pre-study survey variables (Table \ref{tab:participants}) were added as covariates.
Since predictor correlation is a specific case of multicollinearity \cite{alin2010multicollinearity}, pre-study survey variables with significant correlations (Table \ref{tab:popcor}) were excluded as predictors; including participants' age, pre-study motivation to act and motivation to learn.
Participants' ethno-racial backgrounds were not included in the analyses because the sample did not include large enough representative groups to make conclusions. 
Gender was also not included in the regressions, since the sample did not include a large enough non-binary representation, and preliminary analyses did not yield significant differences in responses for participants identifying as women versus men. 

\begin{table}[htbp!]
\caption{Significant moderate or large correlations (i.e., Pearson's $r$) within population variables ($|r| \geq .4$, $p < .05$) \cite{dancey2007statistics}.}
\vspace{-5mm}
\begingroup
\begin{center}
\label{tab:popcor}
\begin{tabular}{ p{0.42\linewidth} p{0.29\linewidth} C{0.03\linewidth} C{0.1\linewidth} }
  & & $r$ & $p$ \\
  \hline
  Age & Degree completion & .67 & < .001\\
  Concern in climate change & Motivation to act & .66 & < .001\\
  Concern in climate change & Motivation to learn & .62 & < .001\\
  Motivation to act & Motivation to learn & .71 & < .001\\
  
  \hline
\end{tabular}
\end{center}
\endgroup
\end{table}

For models where the outcome variable was a repeated measurement, i.e., a follow-up (two weeks later) measurement that was also measured at post-video, since there were only two time points (post-video and follow-up), we chose to model follow-up measurements as differences from the post-video measurements.
In other words, each follow-up measurement's corresponding post-video measurement was added as a predictor; these follow-up measurements include follow-up distress, state compassion, RTS Vid, RTS Qn, concern for climate change, motivation to act, motivation to learn, equality of value of lives, and Mezirow reflection classification.
Nominal predictors (e.g., whether participants completed their bachelor degree, received STEM education, reported themselves as native English speakers) were coded using weighted effects coding to avoid biases due to unequal group sizes (e.g., there were more STEM than BHASE participants) in the model \cite{te_Grotenhuis_2016}. 
On the other hand, continuous (e.g., SOCS-O, SDS) and ordinal predictors (e.g., pre-study concern for climate change, participants' prior knowledge and trust in video) were centered for the regression analysis since this allows for the meaningful interpretation of interaction and main effect terms within interaction models \cite{schielzeth2010simple}.

For each outcome variable, saturated models were first built by including all predictors and covariates, as explained above.
Linear models were built for continuous outcome variables (including post-video and follow-up distress, state compassion, RTS Vid, RTQ Qn, and recall), proportional odds models were built for ordinal outcome variables (including how surprising and emotional participants found the video to be at post-video, and post-video and follow-up concern for climate change, motivation to act, motivation to learn, equality of value of lives, and Mezirow reflection classification), and binomial regression models were built for the binary outcome variable of prosocial behavioral changes.
Insignificant predictors were removed while ensuring reasonable model fit using the likelihood ratio test and variance inflation factor. 
 \cite{gareth2013introduction}.
Outcome variables that were not significantly predicted by condition are reported in Appendix \ref{appendix:rq1}.

\subsubsection{Results}\label{sec:convsrerstime1}~\\
\noindent\textbf{Differences in Post-Video Effects.}
In terms of short term effects, RE/RS participants had a significantly higher odds of reporting greater post-video motivation to learn ($\beta = 1.71, p < .01, OR = 5.54 (95\% CI: 1.71, 19.98)$), even when controlling for covariates such as initial levels of motivation to learn ($\beta = 0.96, p < .001, OR = 2.60 (95\% CI: 1.48, 4.97)$) and level of education ($\beta = -1.03, p < .001, OR = 0.36 (95\% CI: 0.19, 0.63)$).
An example is RS16, who said, ``I will definitely be learning more about this [issue] ... How can I deal with sea level changes? How do you make an impact towards sea levels?'' during post-video.

\begin{figure}[!tbp]
  \centering
    \includegraphics[width=.5\textwidth]{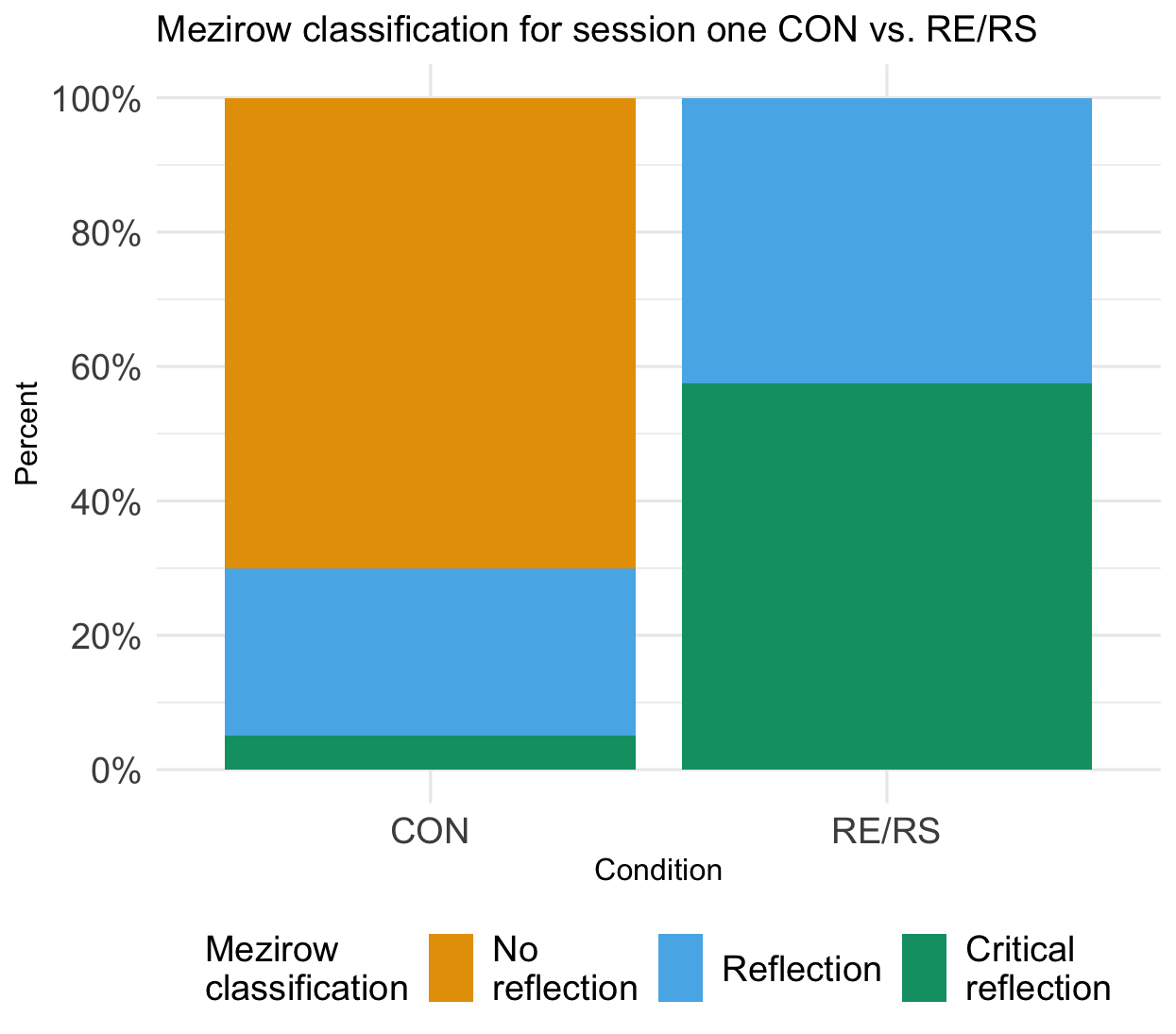}
    \caption{Plot showing level of reflection performed by CON versus RE/RS participants during session one according to qualitative analysis using Mezirow's reflection levels.}
    \Description{The stacked bar chart shows the level of reflection performed by control versus RE or RS participants during session one according to qualitative analysis using Mezirow's reflection levels. More than half of the control participants performed no reflection, while all participants in the RE and RS conditions performed either reflection or critical reflection.}
    \label{fig:mezirowt1}
\end{figure}

Participants who were in the RE/RS conditions had a significantly higher odds of demonstrating deeper reflection (according to the Mezirow reflection classification) in session one ($\beta = 5.46, p < .001, OR = 235.18 (95\% CI: 30.81, 5408.08)$) (Figure \ref{fig:mezirowt1}). E.g., ``I knew that the water levels are rising, but I didn't think about the fact that it would wash the beaches away, and I didn't connect that with the fact that turtles and other animals lay eggs on the beaches, making beaches extremely important for them'' (RS3). This is after controlling for covariates, including whether they studied in STEM programs ($\beta = .62, p < .01, OR = 1.85 (95\% CI: 1.24, 2.96)$) and level of education ($\beta = -.62, p = .04, OR = .54 (95\% CI: .28, .96)$).\\

\begin{figure}[!tbp]
  \centering
    \includegraphics[width=.48\textwidth]{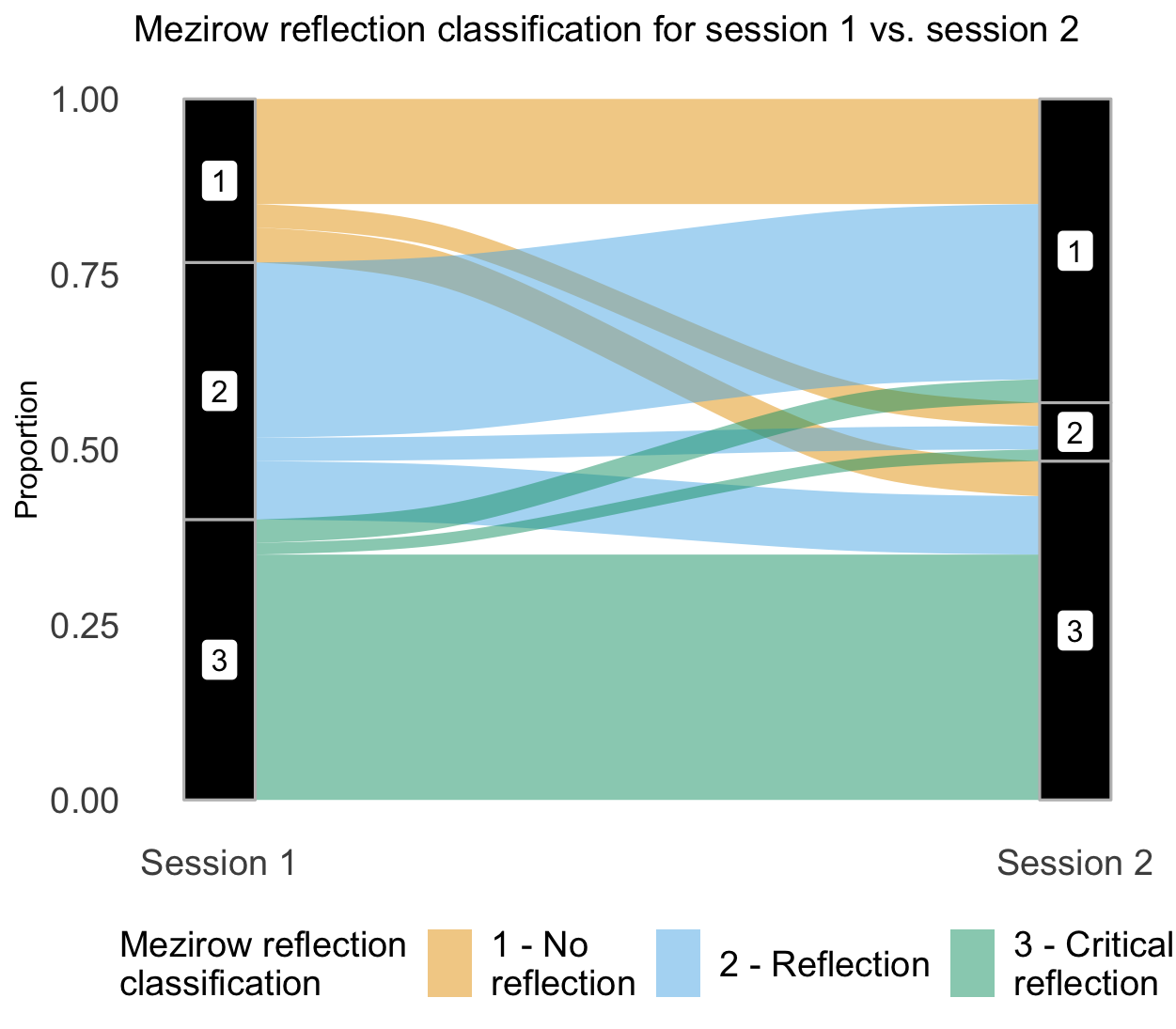}
    \caption{Plot showing the change in participants' reflection level from session one to session two regardless of condition.} 
    \Description{It is a Sankey diagram showing the change in participants' reflection level from session 1 to session 2 regardless of condition. A small portion of participants who performed no reflection in the first session performed either reflection or critical reflection in the second session. Among participants who performed reflection in the first session, about half of them performed no reflection in the second session, and a majority of the other half performed critical reflection in the second session. Most of those who performed critical reflection in the first session also performed critical reflection in the second session.} 
    \label{fig:mezirowt1t2}
\end{figure}

\noindent\textbf{Differences in Follow-up Effects.}
We were also interested in the longer term effects of the reflection processes as measured two weeks later in the follow-up session.  Results show that there is a significant three-way interaction, where RE/RS participants with higher levels of motivation to learn at both pre-study and post-video have higher odds of also having higher follow-up levels of motivation to learn ($\beta = 1.97, p = .05, OR = 7.15 (95\% CI: 1.00, 51.88)$). 
E.g., RE15 reported during follow-up that ``there is a general increase in interest to want to learn more''. 
This is after controlling for prior knowledge as a covariate ($\beta = 0.62, p = .04, OR = 1.86 (95\% CI: 1.04, 3.40)$).

RE/RS participants who performed deeper levels of reflection during session one had significantly higher odds of having performed deeper levels of reflection during the two weeks after session one ($\beta = 3.40, p < .01, OR = 30.07 (95\% CI: 2.43, 489.98)$). As RE5 mentioned, ``watch[ing] the video and discuss[ing] my own actions as part of the study really helped me evaluate what I was seeing, what I was not doing and what's going on in this world.'' 
Regardless of condition, participants who performed deeper levels of reflection during session one also had higher odds, albeit having a smaller effect size ($\beta = .26, p < .001, OR = 1.30 (95\% CI: .25, 7.79)$), as shown in Figure \ref{fig:mezirowt1t2}.
Otherwise, CON participants reported that ``this study didn't have a long-term impact'' (CON20).
Another covariate is how curious participants reported themselves to be ($\beta = -1.00, p = .03, OR = .37 (95\% CI: .13, .93)$).

Finally, RE/RS participants who performed deeper levels of reflection during session one had significantly higher odds of having prosocial behavioral changes after two weeks; $\beta = 2.78, p = .05, OR = 16.05 (95\% CI: 1.23, 478.79)$, see Section \ref{sec:behaviorchange} for related qualitative findings.

\subsection{RQ2: Effects of Emotion-Focused vs Surprise-Focused Reflection Process}
We present below results on the differences between RE and RS participants.
In contrast to CON participants, RE/RS participants were asked pre-video and post-video reflective prompts (Table \ref{tab:reflectqns}) and instructed to tag video moments that were emotional/surprising while watching the video. Post-video reflection were based on two tags that were the most emotional/surprising.\\

\subsubsection{Analysis Methods}\label{sec:result_rq2analysis}
Differences between RE and RS participants were analyzed in two ways: their video-tagging patterns, and their responses (i.e., to the pre-/post-video reflection questions, post-video and follow-up surveys).
Before analyzing differences in RE/RS tagging patterns, the video's content was first split into 10 parts, as shown in Table \ref{tab:tagsections}. Then, RE/RS participants' tags were categorized into their corresponding video parts and compared across conditions.
On the other hand, differences in RE/RS participants' responses were analyzed using RQ1's analysis process.
The only difference was the removal of CON participants from the analysis. Participants' condition was instead dummy coded such that RS participants had a code of 1 with RE participants acting as the reference (coded as 0).
Models with significant condition effects are presented.\\

\subsubsection{Results}
\begin{figure*}[!tbp]
  \centering
  \begin{minipage}[b]{0.47\textwidth}
    \includegraphics[width=\textwidth]{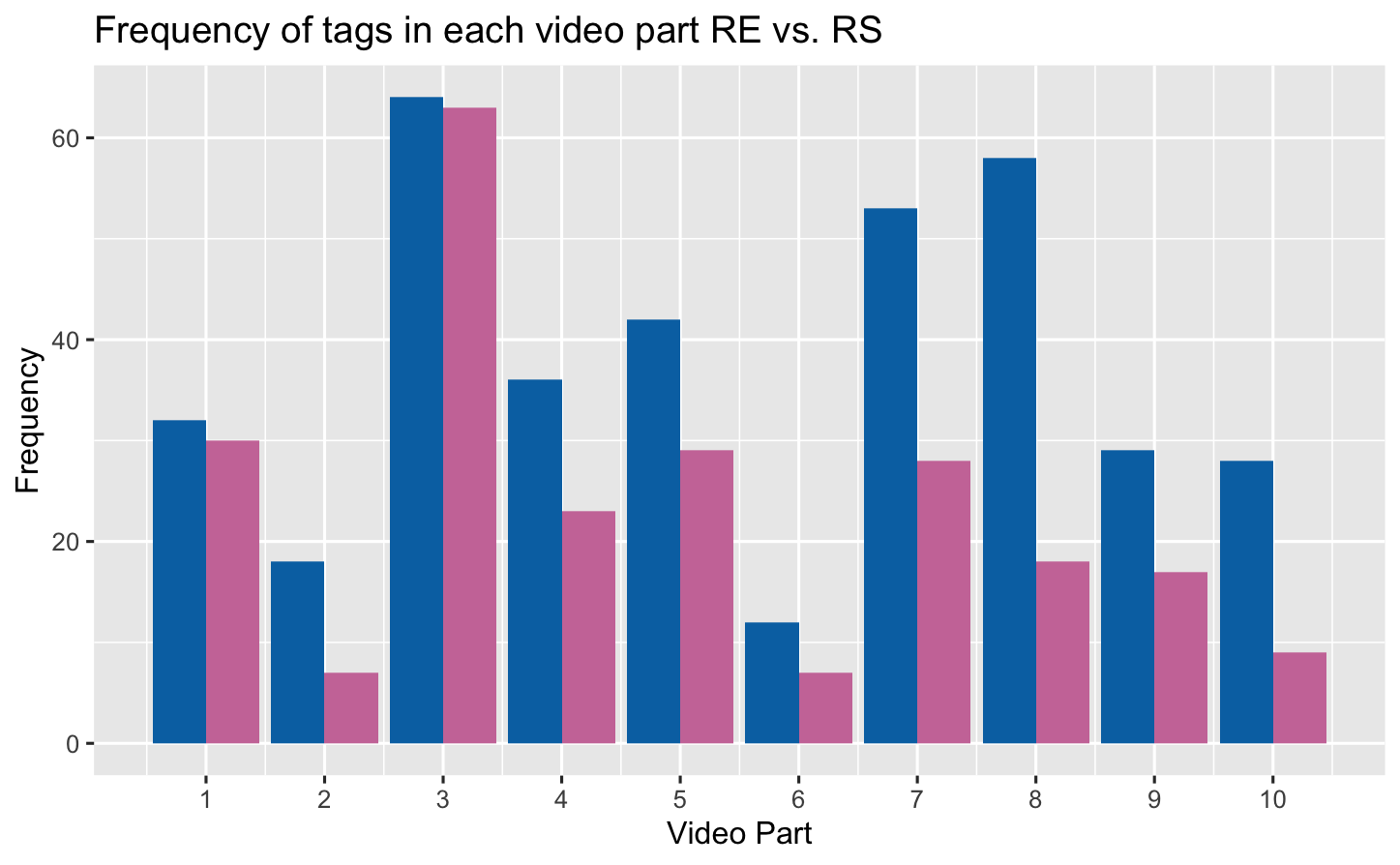}
    \caption{Frequency of tags by video parts, created by RE and RS participants.}
    \Description{This is a grouped bar chart showing the frequency of tags created by RE and RS participants for each of the 10 video parts.}
  \label{fig:alltags}
  \end{minipage}
  \hspace{2mm}
  \begin{minipage}[b]{0.5\textwidth}
    \includegraphics[width=\textwidth]{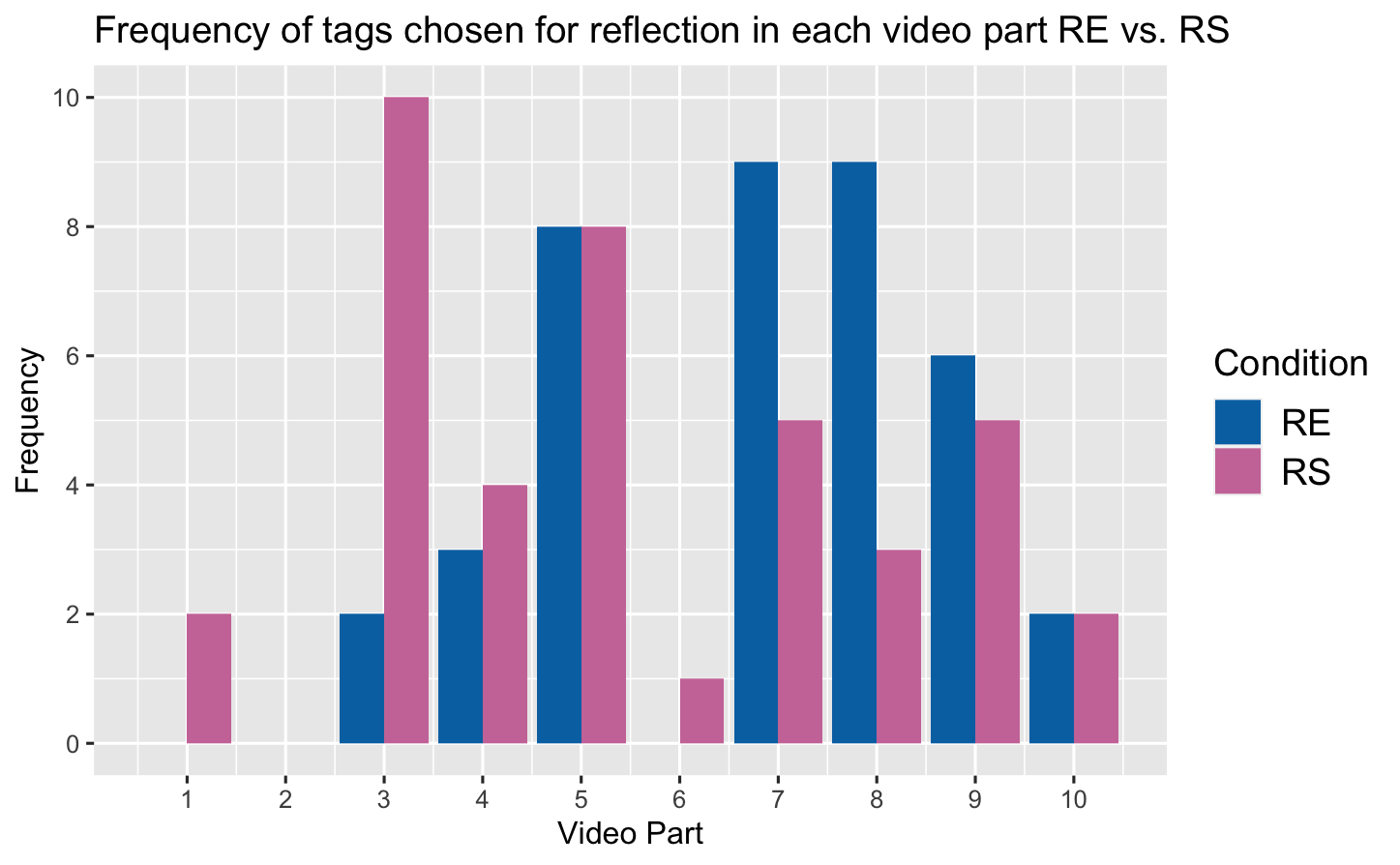}
    \caption{The tags chosen by RE and RS participants to reflect on; RE participants chose two of the most emotional tags, and RS participants chose two of the most surprising tags.}
    \Description{This is a grouped bar chart showing the frequency of tags chosen by RE and RS participants for reflection. RE participants chose two of the most emotional tags, and RS participants chose two of the most surprising tags.}
  \label{fig:twotags}
  \end{minipage}
  
\end{figure*}

~\\
\noindent {\bf Differences in Tagging Pattern.} 
Overall, RE participants created partially significantly more tags than RS participants ($M_{RE}$=18.6$\pm13.8$, $M_{RS}$=11.6$\pm7.94$, $t(30.40) = 1.98$, $p = .06$), as shown in Figure \ref{fig:alltags}.
Interestingly, both RE and RS participants created the most tags in the same part of the video (Part 3) when the video first presents visual evidence of islands that are either sinking or have sunken.
RE participants also found Parts 7 and 8 to be particularly emotional.

\begin{table*}
\caption{Description of the 10 parts of the video used in this study: a 17-minute YouTube documentary created by the 60 Minutes Australia program on the Solomon Islands' rising sea levels \cite{risingseavideo}.}
\begingroup
\begin{center}
\label{tab:tagsections}
\begin{tabular}{ p{0.08\linewidth} c l}
  Part No. & Timestamp Range & Description \\
  \hline
    1 &   0:00 - 1:13 & Video overview \\
    \hline
    2 &  1:13 - 2:38 & Introduction of Dr. Simon Albert's research \\
    \hline
    3 &  2:39 - 5:23 & Visiting islands that have sunken \\
    \hline
    4 &  5:23 - 6:49 & Scientific trends and responding to climate change deniers\\
    \hline
    5 &  6:49 - 8:42 & Interview with Glarus Hibou on the lost of Kale Island \\
    \hline
    6 &  8:42 - 9:33 & Interlude \\
    \hline
    7 &  9:33 - 12:11 & Interview with a Solomon Islands resident on houses getting washed away \\
    \hline
    8 &  12:11 - 13:52 & Interview with Solomon Islands residents on their future\\
    \hline
    9 &  13:53 - 15:43 & Effects on Hawksbill turtles \\
    \hline
    10 &  15:44 - 17:13 & Conclusion and call to action \\
  \hline
\end{tabular}
\end{center}
\endgroup
\end{table*}

Figure \ref{fig:twotags} shows the frequency of the two chosen tags (that the RE and RS participants were asked to reflect on) in each video part.  Although RS participants most frequently chose tags in Part 3, RE participants chose tags in Parts 7 and 8 most often instead.
However, tags in other parts, like Parts 5 and 9, are selected by participants in both RE and RS conditions as being particularly emotional/surprising with approximately equal frequencies. The connection between emotion and surprise was explained by some participants, e.g., the tags ``were factual since it created the emotions, and I was paying attention to it ... my focus was there, and it created the emotional response'' (RE15).

~\\\noindent {\bf Differences in Post-Video Effects.} 
RS participants had higher odds of finding the video more surprising ($\beta = 1.60, p = .02, OR = 4.95 (95\% CI: 1.29, 21.27) $), after controlling for covariates like prior knowledge ($\beta = -1.38, p < .01, OR = 0.25 (95\% CI: 0.09, 0.61)$), whether participants identified as native English speakers ($\beta = -0.96, p = .02, OR = 0.38 (95\% CI: 0.16, 0.86) $), and pre-study level of concern ($\beta = -1.09, p = .02, OR = 0.34 (95\% CI: 0.13, 0.83)$).
This aligns with qualitative responses, e.g., RS13 felt ``extremely shocked'' when learning about Kale Island sinking (Part 5 in Table \ref{tab:tagsections}).\\

\noindent {\bf Differences in Follow-Up Effects.}
RS participants who had higher motivation to learn at pre-study ($ \beta = 1.62, p = .01, OR = 5.07 (95\% CI: 1.44, 20.57)$) had higher odds to have higher follow-up levels of motivation to learn. However, RS participants with higher post-video motivation to learn had lower odds to be more motivated to learn two weeks later ($ \beta = -2.08, p = .04, OR = 0.12 (95\% CI: 0.01, 0.90)$). Covariates included RE/RS participants' SOCS-O scores ($ \beta = 0.11, p < .01, OR = 1.12 (95\% CI: 1.03, 1.23) $) and prior knowledge ($\beta = 1.01, p = .02, OR = 2.74 (95\% CI: 1.14, 7.33)$).  Finally, RS participants scored significantly higher recall than RE participants ($\beta = 0.92, t(37) = 2.32, p = .03$). 

\subsection{RQ3: Correlational Analysis}
RQ3 aims to understand relationships between critical thinking, compassion, prosocial attitudes and behaviors both at post-video and follow-up, regardless of condition.

\subsubsection{Analysis Methods}\label{sec:result_rq3analysis}
Bivariate correlational analysis was used to investigate RQ3. Particularly, we calculated the bivariate correlations between post-video measurements (Sections \ref{sec:time1measure} and \ref{sec:qualcoding}) and also between follow-up measurements (Sections \ref{sec:time2measure} and \ref{sec:qualcoding}).
Below, we present correlations that are both significant and have a moderate ($|r| \geq .4$) or large ($|r| \geq .7$) correlation \cite{dancey2007statistics}.\\

\subsubsection{Results}~~~\\
\noindent {\bf Post-Video Effects.} Regardless of condition, participants' change in concern for climate change was significantly correlated with their change in motivation to act ($r = .72, p < .001$) and motivation to learn ($r = .48, p < .001$). These three measurements were also correlated with RTS Vid and/or RTS Qn (Appendix \ref{appendix:corrtable}).

Participants' post-video state compassion was positively correlated with several other post-video measurements, including change in concern ($r = .44, p < .001$), change in motivation to act ($r = .46, p < .001$), distress ($r = .51, p < .001$), equality of value of lives ($r = .53, p < .001$), as well as how emotional ($r = .49, p < .001$) and surprising ($r = .51, p < .001$) they found the video. Moreover, their compassion was also correlated 
with RTS Vid  - critical reflection ($r = .40, p = .002$), RTS Vid  - reflection ($r = .44, p < .001$) and RTS Qn - reflection ($r = .61, p < .001$).

Participants' post-video Mezirow reflection level was significantly correlated with their post-video motivation to learn ($r = .43, p < .001$).
Other than that, participants' post-video perception of how emotional the video was correlated with their perception of equality of value of lives ($r = .41, p = .001$) and how surprising they found the video ($r = .46, p < .001$).\\

\noindent {\bf Follow-Up Effects.} 
Just like at post-video, participants' change in concern for climate change was significantly correlated with their change in motivation to act ($r = .69, p < .001$) and motivation to learn ($r = .73, p < .001$). Moreover, changes in concern and motivation to act were also correlated with RTS Vid  and RTS Qn (Appendix \ref{appendix:corrtable}).
Participants' follow-up state compassion was correlated with their perception of equality of value of lives ($r = .49, p < .001$).
Interestingly, their session two Mezirow reflection level was negatively correlated with their follow-up state compassion ($r = -.40, p < .01$).

\subsection{Qualitative Evidence of Prosocial Behavioral Changes}\label{sec:behaviorchange}

Participants were asked during the follow-up session about their reflective outcomes, namely, their behavioral and attitudinal changes since the first session. Responses were coded according to Section \ref{sec:qualcoding}, and some of the key findings related to behavioral changes (coded as BR6, Table \ref{tab:qual}) are presented here.

As mentioned in Section \ref{sec:result_resproc}, RE/RS participants with deeper session one reflection demonstrated significantly more prosocial behavioral changes than control participants.  Between the two sessions, a few participants reported using more environmentally-friendly modes of transportation, e.g., ``I usually take Uber ... but now I try to walk when I have time'' (RE2), RE5 ``tried to carpool more'', while RE6 used ``public transportation, cycling rather than using the private car just to decrease the carbon dioxide emissions.''

Others reported adopting more environmentally-friendly consumer behaviors, including buying fewer clothes:
\begin{quote}
    ``I was reading about clothing because making clothes consumes a lot of trees and other materials. For that reason, after the session two weeks ago, I haven’t bought any new clothing, and it’s something I’m going to try to minimize going forward.'' (RE6)
\end{quote}
Furthermore, CON8 ``purchased some vegetarian meat instead of real meat'', CON19 ``tried to purchase fewer things'', RE20 ``tried to buy more locally sourced materials and products'', RE12 ``stopped using plastic bags, bought a lot of recyclable bags to do groceries and asked [their] roommates to use them'' and RS13 ``was about to buy some air freshener'' but ``thought about how it could affect climate change.'' At the end, they ``did not buy it'' and ``started to search for natural alternatives.''

Some participants reported adopting smaller improvements in their daily lives, 
including ``turning off the lights'' when not used (CON19, RS10), 
being ``very careful while segregating the garbage'' (RS16) and learning how to separate recyclables (RE11), having ``stopped pushing the buttons for the automatic door openers'' (RE16) and being ``more aware in the house to not waste electricity and not waste water'' (RE18).

Other types of behavioral changes were also observed. RS2, who ``had the intention of finding a part-time job before [the study],'' was motivated by the first session and decided to find ``a part-time job where [they] could help plant trees'', adding that the job ``does not pay [them] much, but [they] still do it because [they] feel good about it.'' 
Besides, RE5 ``considered switching to a more efficient home heating system from natural gas combustion ... to electrical heat pumps.''
Moreover, RE7 ``thought of planting some new plants and encouraged [their] friends by giving them seeds.''
RS5 ``signed up ... and went to a workshop talking about climate change ... after completing the first part of the study,'' while RE18 decided to pick up garbage that others left at a bus stop because they ``wanted to make some changes.''

\subsection{Other Qualitative Findings}
Regarding RQ1, our qualitative data showed evidence of the importance of the pre-video questions in the RE/RS conditions. In particular, the priming questions elicited participants' knowledge structures, including: 
\begin{itemize}
    \item personal backgrounds and experiences: e.g., ``I come from [a place] where there are a lot of policies aimed at protecting the environment and wildlife'' (RE1)
    \item digital media interaction patterns: e.g., ``reading news articles and even watching documentaries online have shaped my views to view climate change as a big deal and I should be doing something to help prevent it'' (RS3)
    \item current environment-related behaviors: e.g., ``I try to have my reusable water bottle everywhere I go, to limit driving my car and all that stuff as much as I can'' (RS14)
    \item current environment-related beliefs: e.g., ``I think that you do what you can. Not everyone can do everything, so you do what is workable for you'' (RE15)
\end{itemize}
These elicited knowledge structures served as important anchors for their reflections post-video. For instance, when discussing smaller islands disappearing entirely due to sea level rises, RE4 commented ``this is something they discuss all the time as something that could happen in the Caribbean, but we never really see it because we don't frequent our smaller islands.''
Besides that, we also found qualitative evidence of participants reflecting on a few elements of compassion, e.g., recognizing suffering (e.g., ``they are actual human beings, and they are suffering'' (RE18)), feeling empathy (e.g., ``I developed a sort of empathy where I wish I could do something to help the people directly'' (RE4)), and motivation to act (e.g., ``I might try to get more involved in climate change advocacy'' (RE16)).

Regarding RQ2, as discussed above, how surprising RE/RS participants found the video was significantly and positively correlated with how emotional they found the video.
In other words, new information presented in the video often managed to elicit emotions from both RE and RS participants, e.g., ``it was extremely surprising ... a bit sad ... a bit scary'' (RS13). 
However, beyond this similarity, some interesting differences between RE and RS were also observed.
First, RS participants found the video more surprising than RE participants, even though RE participants did not find the video more emotional than RS participants.
Qualitative data supported this; since RS participants were prompted to focus on surprising information, they were more focused on newer aspects of the video (e.g., RS7 felt that it was ``crazy [that] in 10 to 15 years an entire island can be wiped off the map'').
In contrast, RE participants were prompted to focus on an aspect (i.e., emotion) that was already quite salient (e.g., ``I felt sad, but I was also not surprised'' (RE16)) such that further cuing had less effect. 
Finally, RS participants were better at recall during the follow-up session, which may be due to RS participants focusing more on statistics presented during the video, e.g., both RS1 and RS7 distinctly commented on how ``crazy'' it was that the sea level rose by ``15 centimeters'' in the Solomon Islands during post-video.

\section{Discussion}\label{sec:discussion}

\subsection{Key Insights from the Findings}

\subsubsection{RQ1: Effects of Reflection Process.}
Our results show that the reflective conditions successfully made participants verbalize deeper levels of reflection, according to the Mezirow classification, in session one.  
More importantly, only RE/RS participants with deeper Mezirow reflection in session one demonstrated deeper reflection and prosocial behavioral changes two weeks later. 
Regardless of condition, those who performed deeper session one reflection also demonstrated deeper Mezirow reflection two weeks later, albeit to a smaller extent. 
This means the reflection questions were effective at encouraging longer-term critical reflection {\it only} when participants were willing to engage in reflection.  The one RE participant removed from analyses (Section \ref{sec:participants}) was an illustrative example of how the lack of willingness to engage renders the reflection process ineffective. These results align with research showing openness towards reflection as a crucial component of reflection \cite{mamede2004structure,levine2008impact}; ``when we are open to the truth claims put forward by our interlocutors, we can examine critically the prejudices that make up our horizon and potentially `fuse' them with those of our interlocutor'' \cite{cauchon20225}.
Having an open mind in the ``process of communication and a trust in the ability of others to engage likewise'' and be ``less certain of some of our certainties'' is crucial for learning \cite{roberts2011openness}.

The reflection processes also significantly increased participants' post-video motivation to learn.
RE/RS participants with higher pre-study and post-video motivation to learn reported higher increases in motivation to learn two weeks later, showing that the reflection process \textit{avoided} the ceiling effect \cite{van2015scientific}.
In other words, critical reflection was effective at allowing motivated participants to understand their motivation and further increase it.

Finally, reflective conditions' participants did not report significantly higher state compassion, possibly because the reflection questions focused solely on encouraging critical reflection, instead of also aligning with all five components of compassion \cite{Gu2019}. Overall, \textbf{H1} is partially supported.

\subsubsection{RQ2: Emotion-Focused vs. Surprise-Focused Reflection Process.}
Even though the instructions for the two reflection conditions were clearly different---the RE condition asked participants to reflect on scenes that elicited positive/negative emotions versus the RS condition, which asked participants to reflect on scenes that were surprising---the distinction between focusing on emotion vs. information is more blurred than originally expected.
In particular, how surprising RE/RS participants found the video was significantly and positively correlated with how emotional they found the video.
Similarly, the parts that RE and RS participants chose to reflect on did not  significantly differ, partially supporting \textbf{H2}.

However, some interesting differences between RE and RS were also observed.
First, RS participants found the video more surprising than RE participants, even though RE participants did not find the video more emotional than RS participants.
During follow-up, RS participants were better at recall.
This aligns with prior research showing curiosity's ability to encourage engagement with information \cite{vogl2019surprise}; that is, the RS reflection process probed participants to improve their understanding of how newly learned information fitted into their existing knowledge base \cite{dyche2011curiosity}, thereby encouraging a deeper level of processing.

Overall, our findings suggest cognitive-focused, or surprise-focused, reflection to be more beneficial than emotion-focused reflection in terms of engagement with information (i.e., recall) and fostering a motivation to learn.
Although prior research has found emotion- and empathy-based approaches to be effective at encouraging prosocial attitudinal changes \cite{Herrera_2018, batson1997empathy, batson1995information}, we show that cognitive-based approaches, which have been underexplored, could have additional benefits, even in longer-terms.

\subsubsection{RQ3: Relations Between Compassion, Critical Reflection and Other Constructs}
RQ3 aims to understand, regardless of condition, the relationships between critical thinking, compassion, prosocial attitudes and behaviors both at post-video and follow-up.
Our results partially support \textbf{H3.1}; that is, state compassion does correlate positively with reflection (RTS Vid and Qn) and critical reflection (RTS Vid) at post-video, but not at follow-up.
Similarly, we found trait compassion to positively predict both reflection (RTS Vid and Qn) and critical reflection (RTS Qn), but only at post-video.

Those who reported deeper levels of critical reflection reported higher state compassion after digital media consumption; in other words, critical reflection could elicit compassion after digital media consumption, even though the specific reflection questions used in this study did not lead to significantly higher state compassion.
However, follow-up correlations were different from post-video.
Participants' session two Mezirow classification was negatively correlated with follow-up state compassion. Since compassion could be framed translating empathy and prosocial motivations into prosocial behaviors \cite{jeste2020wiser}, the two weeks between the sessions could have provided participants time to complete this translation process. 
A fade away was also observed in the positive correlation between state compassion, concern and motivation to act, which only existed at post-video, hence partially supporting \textbf{H3.2}.

Finally, \textbf{H3.3} is not supported. Post-video state compassion was positively correlated with post-video distress.
This contrasts existing research that found trait compassion for others (measured using the SOCS-O scale) to be significantly and negatively correlated to trait personal distress \cite{Gu2019}, and is most likely due to limitations with the 8-item state compassion measurement.

\subsection{Design and Technological Implications}

\subsubsection{Reflection for Nurturing Compassion}
Although the use of reflective prompts is a common reflection methodology~\cite{fleck2010reflecting}, this work explores its use for a novel target construct: compassion.
Unlike empathy, interventions targeting compassion implies targeting components beyond  empathic feelings~(Table \ref{tab:compassionelem}).
Our study showed that reflection prompts in the context of consuming a piece of digital media could be effective at converting feelings of empathy to prosocial attitudes and behaviors. However, we did not find evidence that the reflection process supported two of the five components of compassion, specifically, the understanding of common humanity and the toleration of negative emotions that resulted from empathy (i.e., resilience).
One possible interpretation here is that to nurture compassion, we must design into the reflection process a way to \textit{facilitate the understanding of multiple perspective towards achieving common humanity}.  There are multiple methodologies for achieving this---for example, Fleck and Fitzpatrick \cite{fleck2010reflecting} proposed the use of technologies to support dialogic reflection (R2) by enabling users to see from multiple perspectives; and perspective-taking technologies for eliciting empathy commonly do so by providing users with perspectives of specific targets (e.g., children, refugees, victims of bullying) \cite{martingano2021virtual}.  Nurturing compassion for broader targets might require deep reflections across multiple perspectives simultaneously. 
For instance, tools could be built to support critical reflection across multiple pieces of related digital media. 
In the case of sea level rises due to climate change, the tool could encourage users to search for digital media on i) the effects of sea level rises across different geographical regions, ii) its effects on both current and future generations, and iii) efforts by various locations and levels (e.g., governments, communities) related to dealing with this issue.
This not only aligns with the elicitation of solidarity discussed by Varma \cite{varma2019empathy}, but also helps to motivate prosociality through humanization  and knowledge of how others are helping with an issue.

\subsubsection{Encouraging Longer-Term Critical Reflection}
As noted by Kember \cite{kember2008reflective}, critical reflection and perspective transformation (R3 and R4 in Fleck and Fitzpatrick's model \cite{fleck2010reflecting}) likely take time. As such, supporting real positive changes in belief systems and resulting behaviors requires technologies that support not just one-time or short-term reflection, but longer-term critical reflection. 
Importantly, like many other contexts for reflection (e.g., sleep data \cite{bentvelzen2022revisiting}), digital media consumption is a long-term activity.
While technologies supporting long-term reflection have been explored \cite{bentvelzen2022revisiting}, there may still be gaps in understanding the unique dynamics of multi-session and long-term reflection.
For example, we found that engaging participants in deeper reflection in the first session is crucial for continued deep reflection two weeks later. 
As such, we encourage designers to \textit{maximize the chances for participants to engage in deeper reflection during the first session}. 
Towards this, special attention should be paid to personalizing the reflection process, especially during the first session, to each person's prior experience/expertise and their willingness to engage in reflection.
Although using technology to support personalized reflection is nothing new \cite{cho2022reflection}, these two aspects have not been properly investigated.
Thinking of the reflection process as a framework, customizable components could include (i) the digital media to reflect upon, (ii) the reflection method (e.g., question-asking, emotion-focused or cognition-focused) and (iii) the sequence of reflection-related activities (e.g., pre-video questions before tagging video).
Technology can be designed to support these customized reflection procedures. 
For instance, different media could be chosen according to a person's preferences when engaging in reflection, e.g., ``some individuals may prefer climate change messages framed without an emotional narrative at all'' \cite{Ettinger2021}.
Priming (i.e., pre-video) questions can be used to promote engagement with reflection, as our study has found.

\subsection{Limitations}
Even though the second and third authors were blind to the research questions when helping with the qualitative coding process, it was unfortunately not possible to be completely blind to which condition participants belonged to while coding, owing to the obvious difference in the participants' transcript lengths (i.e., reflective conditions' participants had much longer transcripts). As such, this might have been a source of experimental bias. The sample was also not representative for all ethno-racial and gender groups.
Moreover, the use of interview questions during the follow-up session asking participants to recall their behavioral changes in the past two weeks was not ideal. Other retrospective methods could have been better at minimizing participant bias \cite{brito2017demonstrating} and retrospection bias \cite{russell2014looking}.

The present study only explored its research questions in one single context--climate change, and only using one piece of digital media within this context. Although qualitative coding was used to complement self-reported measures on critical reflection, a limitation is the significant use of self-reported measures for other constructs.
Furthermore, our findings were limited to uncovering only positive associations, but not causal relationships, between critical reflection and compassion. Further work could also be done in developing an appropriate tool for measuring state compassion.

\section{Conclusion}
The present study aimed to explore, in the context of climate change digital media (i.e., a publicly accessible documentary) consumption, the use of critical reflection via question-asking to elicit feelings of compassion and its constituent component, namely, a motivation to learn and act towards mitigating climate change. 
Since compassion consists of both emotional and cognitive components, we investigated two reflection processes, one that focused on the identification of emotional scenes, and another that focused on identifying surprising facts/scenes from the documentary.
Participants were measured for their reflection level, compassion, prosocial attitudes and behaviors right after the session, as well as two weeks later.
Analyses found positive correlations between critical reflection and state compassion at the end of the session.
The reflective conditions led to deeper reflection, more prosocial behavioral changes and higher motivation to learn even after two weeks, with a willingness to engage in reflection playing an important role.
Between the two reflective conditions, the surprise-focused reflection process yielded 
better recall two weeks later.
Furthermore, this work points out the importance of designing reflection processes that target both compassion's components and long-term critical reflection, and are personalized to an individual's prior experiences and willingness to engage in reflection.

\begin{acks}
We thank all participants for their contributions, lab colleagues for their helpful feedback on the draft and acknowledge the funding from NSERC Discovery Grant RGPIN-2015-04543 for making this work possible.
\end{acks}

\bibliographystyle{ACM-Reference-Format}
\bibliography{main}


\appendix

\newpage
\section{Participants' Ethno-racial Background}\label{appendix:ethnoracial}

\begin{table}[!h]
\caption{Participants' ethno-racial background did not demonstrate any significant simple effects across conditions ($\chi^2 (14, N= 60)$ = 10.11, \textit{p} = .75). Only one participant identified themselves as having more than one ethno-racial background, and this participant was removed during the analysis of the simple effect of participants' ethno-racial background. }
\begingroup
\small
\begin{center}
\begin{tabular}{ p{0.23\linewidth} C{0.15\linewidth} C{0.15\linewidth} C{0.15\linewidth} }
   & Control (\textit{n}=20) & Reflect-Emotion (\textit{n}=20) & Reflect-Surprise (\textit{n}=20)  \\
  \hline
 Black & 1 & 2 & 0 \\
 \hline
 East Asian & 10 & 6 & 5 \\
 \hline
 Indigenous & 0 & 1 & 1\\
 \hline
 Latino & 0 & 2 & 1\\
 \hline
 Middle Eastern & 1 & 1 & 1  \\
 \hline
 South Asian & 5 & 8 & 7 \\
 \hline
 Southeast Asian & 0 & 0 & 2  \\
 \hline
 White & 2 & 3 & 3 \\
 \hline
\end{tabular}
\end{center}
\endgroup
\end{table}

\section{Supplementary Correlation Information}\label{appendix:corrtable}

\subsection{Post-video}
\begin{table}[!h]
\caption{Post-video supplementary correlation information.}
\begingroup
\small
\begin{center}
\begin{tabular}{ p{0.3\linewidth} c c c c c c  }
   & 1 & 2 & 3 & 4 & 5 & 6 \\
  \hline
 1. Change in Concern & - & &&&& \\
 \hline
 2. Change in Motivation to act & $.72^{***}$ & - &  &  &  & \\
 \hline
 3. Change in Motivation to learn & $.48^{***}$ & $.67^{***}$ & - &&& \\
 \hline
 4. RTS Vid - Reflection & $.26^{*}$ & $.28^{*}$ & $.35^{*}$ & - && \\
 \hline
 5. RTS Vid - Critical Reflection & $.48^{*}$ & $.51^{***}$ &.39 & .28 & - & \\
 \hline
 6. RTS Qn - Reflection & .35 & $.41^{**}$ & $.45^{***}$ & $.64^{***}$ & $.51^{***}$ & - \\
 \hline
 7. RTS Qn - Critical Reflection & .30 & $.47^{***}$ & $.51^{***}$ & $.36^{**}$ & $.80^{***}$ & $.61^{***}$ \\
 \hline
\end{tabular}
\end{center}
\endgroup
\textit{Note:} 
  $^*p<.05.^{**}p<.01.^{***}p<.001$
\end{table}

\newpage
\subsection{Follow-up}
\begin{table}[hbt!]
\caption{Follow-up supplementary correlation information.}
\begingroup
\small
\begin{center}
\begin{tabular}{ p{0.3\linewidth} c c c c c c  }
   & 1 & 2 & 3 & 4 & 5 & 6 \\
  \hline
 1. Change in Concern & - &&&&& \\
 \hline
 2. Change in Motivation to act & $.69^{***}$ & - &&&& \\
 \hline
 3. Change in Motivation to learn & $.73^{***}$ & $.79^{***}$ & - &&& \\
 \hline
 4. RTS Vid - Reflection & $.32^{*}$ & $.35^{**}$ & $.26^{*}$ & - && \\
 \hline
 5. RTS Vid - Critical Reflection & $.49^{***}$ & $.44^{***}$ & $.32^{*}$ & $.32^{*}$ & - & \\
 \hline
 6. RTS Qn - Reflection & $.35^{**}$ & $.32^{*}$ & .25 & $.81^{***}$ & $.37^{**}$ & - \\
 \hline
 7. RTS Qn - Critical Reflection & $.47^{***}$ & $.38^{**}$ & $.31^{*}$ & $.33^{**}$ & $.86^{***}$ & $.44^{***}$ \\
 \hline
\end{tabular}
\end{center}
\endgroup
\textit{Note:} 
  $^*p<.05.^{**}p<.01.^{***}p<.001$
\end{table}

\section{Supplementary RQ1 Analysis}\label{appendix:rq1}

\subsection{Post-video}

\textbf{Distress.} Participants' SDS scores significantly increases their post-video distress ($ \beta = 1.21, t(55) = 2.02, p = .05$).

\noindent\textbf{State compassion.} Participants with higher SOCS-O scores ($ \beta = 0.27, t(53) = 4.26, p < .001$) reported significantly higher post-video state compassion. 

\noindent\textbf{RTS Vid - Reflection.}
No significant differences were found, i.e., the final model was an intercept-only model.

\noindent\textbf{RTS Vid - Critical Reflection.}
Participants who had higher SOCS-O scores reported more critical reflection due to the video ($ \beta = 0.04, t(52) = 2.86, p < .01$). Conversely, those with more prior knowledge ($ \beta = -0.38, t(52) = -2.87, p < .01$) or had higher pre-study concern ($ \beta = -0.47, t(52) = -2.99, p < .01$) reported lower critical reflection due to the video. 

\noindent\textbf{RTS Qn - Reflection.}
Participants with higher SOCS-O scores reported slightly higher reflection due to the open-ended questions ($ \beta = 0.03, t(54) = 3.25, p < .01$), while those with more prior knowledge reported less reflection ($ \beta = -0.26, t(54) = -2.68, p = .01$).

\noindent\textbf{RTS Qn - Critical Reflection.}
Participants with higher SOCS-O scores reported slightly higher critical reflection due to the open-ended questions ($ \beta = 0.04, t(52) = 2.86, p < .01$), while those with higher prior knowledge ($ \beta = -0.38, t(52) = -2.87, p < .01$) or pre-study concern ($ \beta = -0.47, t(52) = -2.99, p < .01$) reported less critical reflection.

\noindent\textbf{Concern for climate change.} 
Participants who reported themselves to be more emotional as a person had higher odds to report a higher post-video level of concern, by a factor of 2.43 ($\beta = 0.89, p = .02, OR = 2.43 (95\% CI: 1.16, 5.38)$).
Conversely, those who were either English native speakers ($\beta = -0.79 , p = .02, OR = .45 (95\% CI: 0.22, 0.89)$) or reported a higher level of prior knowledge ($\beta = -1.15, p < .001, OR = .32 (95\% CI: 0.16, 0.60)$) had lower odds to report having more concern.

\noindent\textbf{Motivation to act.}
Participants who identified as native English speakers ($\beta = -0.88, p = .01, OR = 0.42 (95\% CI: 0.20, 0.83)$) or had higher prior knowledge ($\beta = -0.92, p < .01, OR = 0.40 (95\% CI: 0.20, 0.74)$) had lower odds of having a higher level of motivation to act.

\noindent\textbf{Equality of value of lives.}
Participants who had higher levels of pre-study concern had higher odds to perceive the value of the lives of their loved ones and those of the Solomon Islanders as equal ($\beta = 1.08, p < .01, OR = 2.95 (95\% CI: 1.34, 7.19) $).

\noindent\textbf{How emotional is the video?}
Participants who had more trust in the video ($\beta = 1.17, p < .01, OR = 3.21 (95\% CI: 1.34, 8.15)$) had higher odds of being emotional of the video.

\noindent\textbf{How surprising is the video?}
Participants with higher SDS scores ($\beta = 0.36, p = .01, OR = 1.44 (95\% CI: 1.08, 1.96)$) or trusted the video more ($\beta = 0.92, p = .04, OR = 2.50 (95\% CI: 1.03, 6.33)$) had higher odds of perceiving the video as more surprising. In contrast, those who completed their bachelor's degree ($\beta = -0.53, p = .04, OR = 0.59 (95\% CI: 0.34, 0.99)$), had more prior knowledge ($\beta = -0.98, p < .001, OR = 0.38 (95\% CI: 0.21, 0.66)$), or had higher pre-study concern ($\beta = -0.89, p = .01, OR = 0.41 (95\% CI: 0.19, 0.82)$) had lower odds instead.

\subsection{Follow-up}

\textbf{Distress.} Participants who were native English speakers ($ \beta = -4.31, t(55) = -2.47, p = .02$) reported significantly less distress.

\noindent\textbf{State compassion.} Participants with higher post-video state compassion scores also reported higher follow-up state compassion with a small effect size ($ \beta = 0.28, t(53) = 2.11, p = .04$).

\noindent\textbf{RTS Vid - Reflection.}
Participants who reported having been through STEM university education reported less reflection  ($ \beta = -0.11, t(53) = -2.01, p = .05$).

\noindent\textbf{RTS Vid - Critical Reflection.}
Participants who identified as native English speakers ($ \beta = -0.41, t(53) = -2.93, p < .01$) reported lower critical reflection due to the video.

\noindent\textbf{RTS Qn - Reflection.}
Participants who reported themselves to be more curious as a person also reported having more reflection due to the open-ended questions ($ \beta = 0.34, t(54) = 2.49, p = .02$), while those who studied a STEM program reported lower reflection ($ \beta = -0.13, t(54) = -2.05, p = .05$).

\noindent\textbf{RTS Qn - Critical Reflection.}
Participants who identified as native English speakers reported less critical reflection due to the open-ended questions ($ \beta = -0.48, t(55) = -2.98, p < .01$).

\noindent\textbf{Recall.} 
No significant differences were found.

\noindent\textbf{Concern for climate change.} 
Participants who reported themselves to have higher pre-study concern about climate change had lower odds to report a higher follow-up level of concern, by a factor of .51 ($\beta = -0.68, p = .04, OR = 0.51 (95\% CI: 0.25, 0.98)$).

\noindent\textbf{Motivation to act.}
Neither of participants' condition, population variables or post-video measurements significantly influenced their follow-up motivation to act, i.e., the final model was an intercept-only model.

\noindent\textbf{Equality of value of lives.}
Participants who reported as having been through STEM university education had lower odds of perceiving the value of the lives of their loved ones and the Solomon Islanders as equal ($\beta = -0.38, p = .03, OR = 0.69 (95\% CI: 0.46, 0.97)$).


\end{document}